\documentclass[twocolumn, numberedappendix]{emulateapj}
\usepackage{amssymb,amsmath}
\usepackage{graphics,color,rotating}
\usepackage{ hyperref }

\begin{document}

\title{The Intrinsic Neptune Trojan Orbit Distribution:\\ Implications for the Primordial Disk and Planet Migration}

\author{\bf Alex H. Parker$^{1\dagger}$ }
\email{$\dagger$ alexharrisonparker@gmail.com}
\affil{\emph{ $^{1}$Department of Astronomy, University of California at Berkeley} }

\shortauthors{Parker}
\shorttitle{The Neptune Trojan Orbit Distribution}

\begin{abstract}

The present-day orbit distribution of the Neptune Trojans is a powerful probe of the dynamical environment of the outer solar system during the late stages of planet migration. In this work, I conservatively debias the inclination, eccentricity, and libration amplitude distributions of the Neptune Trojans by reducing \emph{a priori} unknown discovery and follow-up survey properties to nuisance parameters and using a likelihood-free Bayesian rejection sampler for parameter estimation.  Using this \emph{survey-agnostic} approach, I confirm that the Neptune Trojans are a dynamically excited population: at $>$95\% confidence, the Neptune Trojans' inclination width must be  $\sigma_i > 11^\circ$. For comparison and motivation purposes, I also model the Jupiter Trojan orbit distributions in the same basis and produce new estimates of their parameters (Jupiter Trojan $\sigma_i=14.4^\circ \pm 0.5^\circ$, $\sigma_{L11} = 11.8^\circ \pm 0.5^\circ$, and $\sigma_e = 0.061\pm 0.002$). The debiased inclination, libration amplitude, and eccentricity distributions of the Neptune Trojans are nominally very similar to those of the Jupiter Trojans. I use these new constraints to inform a suite of simulations of Neptune Trojan capture by an eccentric, rapidly-migrating Neptune from an initially dynamically-hot disk. These simulations demonstrate that if migration and eccentricity-damping timescales were short ($\tau_a \lesssim 10$ Myr, $\tau_e \lesssim 1$ Myr), the disk that Neptune migrated into \emph{must} have been pre-heated (prior to Neptune's appearance) to a width comparable to the Neptune Trojans' extant width to produce a captured population with an inclination distribution width consistent with that of the observed population.

\end{abstract}

\maketitle

\section{Introduction}

A small sample of Neptune Trojans has been accumulated by a variety of surveys; however, inferences drawn from this sample about the intrinsic distributions of Neptune Trojan orbital properties have been limited and generally qualitative. The challenge inherent in extracting meaningful information from this sample is accurately determining the properties of the surveys that discovered them, and the properties of all surveys that, while sensitive to Neptune Trojans, did not discover any. In this work, I treat these unknown survey properties as nuisance parameters, and marginalize over them to extract as much useful information about the intrinsic orbital distributions of the Neptune Trojans as possible.

Of particular dynamical interest are the inclination, eccentricity, and libration amplitude distributions. These distributions encode information about the formation mechanism (in-situ formation, chaotic capture, or other processes) and post-formation evolution. Several Neptune Trojans have remarkably high inclinations ($\sim25^\circ-30^\circ$), even though surveys have by-and-large targeted fields near the Ecliptic where objects on inclined orbits spend relatively little time. Previous works have noted that this qualitatively indicates the existence of a large, poorly-sampled high-inclination Neptune Trojan population (Sheppard \& Trujillo 2006, Sheppard \& Trujillo 2010a).

In this work, I simultaneously consider the inclination, eccentricity, and libration amplitude distributions, generate synthetic populations of Neptune Trojans defined by these distributions, then pass these synthetic populations through ``coverage functions:'' simplified observational filters that are treated as independent functions of heliocentric ecliptic latitude $\beta$, and heliocentric longitudinal separation from the Trojan libration centers $\lambda'$ (libration centers located roughly $\pm60^\circ$ from Neptune), and inclination. The properties of these observational coverage functions are then marginalized over, effectively marginalizing the unknown properties of the surveys which discovered the Trojans.  Like the Jupiter Trojans and other trans-Neptunian populations, the inclination distribution is modeled as a Brown's distribution ( $p(i) \propto \sin(i) \exp( -i^2 / 2 \sigma^2$ ). The libration amplitude $L_{11}$ and eccentricity $e$ distributions are both modeled as Rayleigh distributions, motivated by the distributions of the Jupiter Trojans. The inclination, libration amplitude, and eccentricity distributions are all truncated at upper limits derived from stability constraints, requiring appropriate corrections to their proposal volume and probability density functions. All statistical analysis is performed in the conceptually simple yet analytically powerful ``Approximate Bayesian Computation'' framework, described in section \ref{ABC}.

\section{Sample}\label{Sample}

\begin{table}[t]
\centering
\begin{tabular}{  l r r r r  }
\multicolumn{5}{c}{\bf Table 1: Adopted Neptune Trojan Properties}\\
\hline
Name & $i^a$ & $L_{11}^b$ & $| \beta |^d$ & $ \lambda'$$^c$ \\
\hline
2001 QR$_{322}$ & 1.3$^\circ$   & 25.5$^{+0.4}_{-0.8}$$^\circ$   &  0.57$^\circ$   &  10.46$^\circ$ \\
2004 UP$_{10}$  & 1.4$^\circ$   & 10.8$^{+1.0}_{-0.3}$$^\circ$   &  0.73$^\circ$   &  10.24$^\circ$ \\
2005 TN$_{53}$  & 25.0$^\circ$  & 8.7$^{+0.3}_{-0.5}$$^\circ$    &  0.62$^\circ$   &  8.51$^\circ$  \\
2005 TO$_{74}$  & 5.2$^\circ$   & 9.2$^{+0.2}_{-0.5}$$^\circ$    &  1.62$^\circ$   &  9.12$^\circ$  \\
2006 RJ$_{103}$ & 8.2$^\circ$   & 6.3$^{+0.1}_{-0.3}$$^\circ$    &  7.99$^\circ$   &  0.58$^\circ$  \\
2007 VL$_{305}$ & 28.1$^\circ$  & 14.2$^{+0.03}_{-0.10}$$^\circ$ &  11.25$^\circ$  &  9.44$^\circ$  \\
2008 LC$_{18}$  & 27.6$^\circ$  & 16.4$^{+1.3}_{-1.1}$$^\circ$   &  2.80$^\circ$   &  0.84$^\circ$  \\
2011 HM$_{102}$ & 29.4$^\circ$  &  9.8$^{+0.4}_{-0.4}$$^\circ$   &  2.60$^\circ$   &  7.72$^\circ$  \\
\hline
\end{tabular}\\
{\footnotesize $^a$: J2000 ecliptic inclination. $^b$: Half-peak RMS libration amplitude and $1\sigma$ uncertainty. $^c$: Absolute value of J2000 heliocentric ecliptic longitude separation of object and nominal Trojan center; see Eqn.  \ref{Lprime}. $^d$: J2000  Heliocentric ecliptic latitude.}
\end{table}

The Minor Planet Center (MPC) lists nine Neptune Trojans (six L4, three L5), but one of the L5 Trojans is unstable and likely a recently captured Centaur (Gladman et al. 2012, Horner et al. 2012). This object is therefore not considered to be reflective of the intrinsic inclination distribution of the (putatively primordial) Neptune Trojans. With the addition of the newly discovered L5 Trojan 2011 HM$_{102}$ (Parker et al. 2013), the 8 known long-term stable Trojans have ecliptic inclinations ranging from $1.3^\circ$ to $29.4^\circ$, and heliocentric ecliptic latitudes at discovery ranging in amplitude from $0.6^\circ$ to $11.7^\circ$. Figure 1 illustrates these properties, and it is clear that Trojans have generally been discovered at latitudes significantly lower than their inclinations, even though an object spends roughly 50\% of their time at latitudes greater than 70\% of their inclination. Only the object 2006 RJ$_{103}$ was higher than its median latitude at the time of discovery. This indicates that it is likely that most surveys that discovered Neptune Trojans targeted the ecliptic, and were therefore strongly biased toward detecting low inclination objects, and yet discovered a surplus of high-inclination objects. The larger number of known L4 Neptune Trojans compared to L5 is likely an artifact of the L5 cloud being more poorly surveyed due to its current proximity to the Galactic plane.

These Neptune Trojans were discovered by a variety of surveys, performed at a variety of facilities and under varying conditions, and normally would not represent a sample from which estimating an intrinsic, debiased orbit distribution would be statistically advisable. However, using appropriate statistical care, we can make a conservative estimate of the range of plausible properties of the Neptune Trojan orbital distribution by marginalizing over the plausible volume of the unknown characteristics of all discovery surveys. This {\it survey-agnostic} approach can conceivably be applied to other populations, and since it is performed in a Bayesian framework, the outcomes can be meaningfully combined with results from large, monolithic, well-characterized surveys such as DES (eg., Gulbis et al. 2010), CFEPS (eg., Petit et al. 2011) and the ongoing \emph{Outer Solar System Origins Survey}\footnote{CFHT Large Program proposal: \url{http://cfht.hawaii.edu/en/science/LP_13_16/OSSOS.pdf}}. Because of their small sample size and currently poorly-characterized orbit distributions, I consider the Neptune Trojans a useful demonstration population.

The libration amplitudes listed in Table 1 were generated with the same technique as Parker et al. (2013). Each object's motion was integrated with \textit{mercury6} (Chambers 1999) in the presence of the giant planets for 1 Myr. 100 clones of each object were integrated, with initial state vectors centered on the JPL Horizons solution, perturbed to populate the Cartesian uncertainty manifold generated by fitting all ground-based observations of each object with the \textit{fit\_radec} and \textit{abg\_to\_xyz} routines developed in association with Bernstein \& Kushalani (2000). Libration amplitudes for each clone were measured by assuming that libration is sinusoidal and deriving the sinusoidal half-amplitude from the RMS of the $n$ samples of the resonant angle over the entire 1 Myr integration:

\begin{equation}
L_{\mbox{fit}} = \left(\frac{2}{n} \sum\limits_{i=1}^n ( \phi_i - \langle \phi \rangle )^2 \right )^{\frac{1}{2}}, 
\end{equation}

\noindent which produces the more appropriate half-amplitude for scaling a sinusoidal model than the usual peak-to-peak definition of $L = \frac{1}{2}[ \max(\phi_i) - \min(\phi_i) ]$. The RMS produces a value that better reflects the mean libration behavior, while defining the amplitude from peak to peak is sensitive to large, single-cycle excursions of the resonant argument. As such, the RMS-defined amplitude is always smaller than the peak-to-peak definition.

\begin{figure}[t]
\centering
\includegraphics[width=0.55\textwidth]{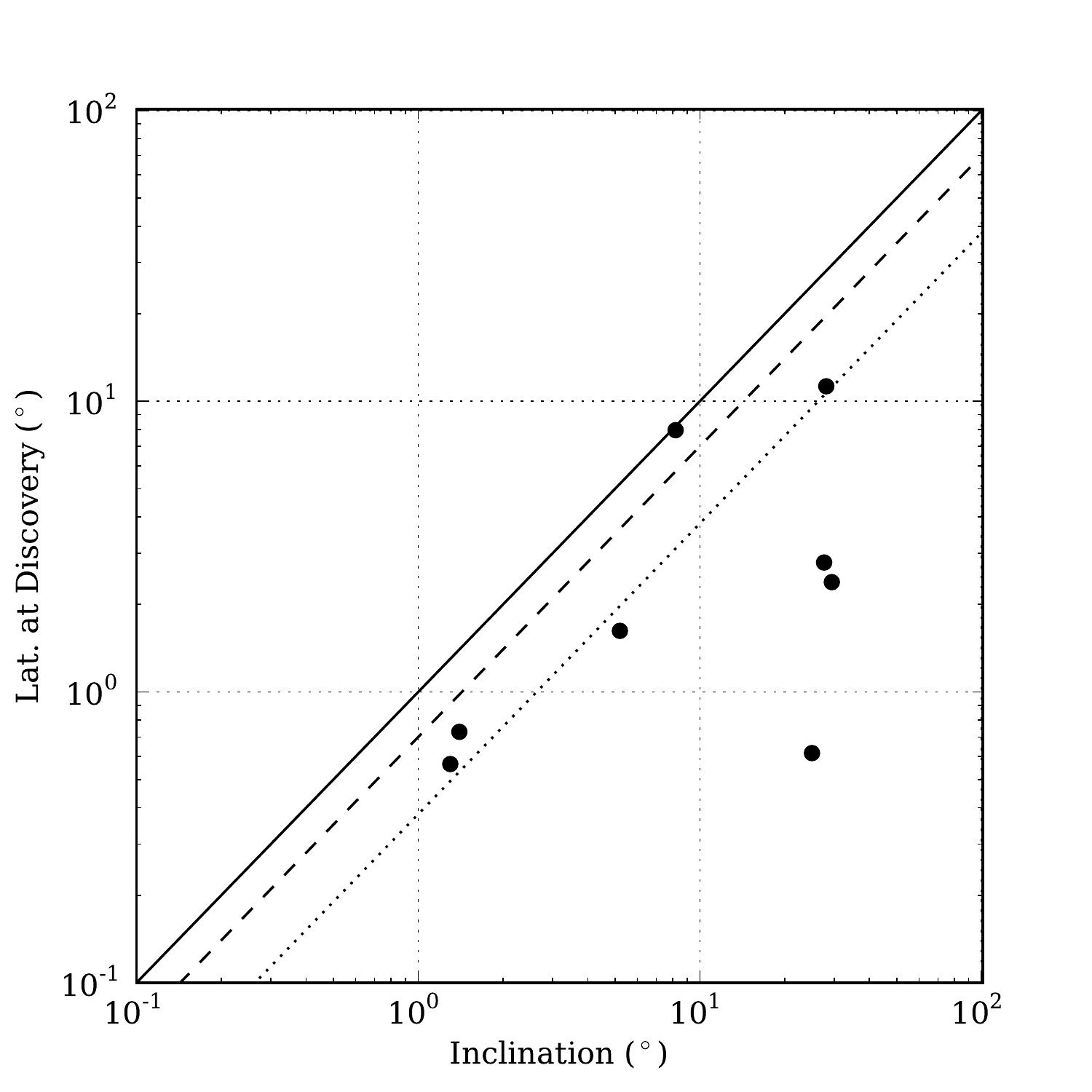}
\caption{Heliocentric ecliptic latitude at discovery vs. inclination for known long-lived Neptune Trojans. Solid line indicates maximum possible latitude achievable for a given inclination, dashed line indicates an object's median latitude for a given inclination, and dotted line indicates an object's lower-quartile latitude for a given inclination. All but one object falls below the median, and half the sample falls in the lowest quartile.}
\label{DiscoveryCirc}
\end{figure}

\section{Approximate Bayesian Computation}\label{ABC}

For all parameter estimation in this work I utilize a likelihood-free rejection sampler --- specifically, the ``Approximate Bayesian Computation'' rejection (ABCr) scheme first presented in Pritchard et al. (1999). Approximate Bayesian Computation (ABC) is conceptually simple but statistically powerful, and has the primary advantage of not requiring the computation of any true likelihood value. I briefly outline this approach below, and refine its description later as merited by the specifics of each application. The literature on ABC is well-developed, with much more sophisticated methods available than those employed here; for a recent review see Marin et al. (2011).

In the context of this work, I utilize ABCr in order to approximate the posterior probability density functions of a set of parameters $\theta$ which define the properties of a model $M$ that describes an observed population of minor planets. Given a metric of similarity $\chi( A, B )$ that reflects the relative similarity of some set of properties of an observed sample $A$ and a synthetic sample $B$, where the synthetic set is generated by a model $M(\theta)$, and where the parameters $\theta$ of $M$ are drawn from prior distribution $\Gamma(\theta)$:

\begin{enumerate}
\item Propose a new set of model parameters $\theta_p$, drawing from the prior distribution $\Gamma$ for each parameter.
\item Generate a synthetic sample of observations $B$ from model $M$ with parameters $\theta_p$.
\item Generate the similarity metric $\chi$ for this synthetic sample compared to the observed sample.
\item If similarity metric $\chi$ is less than some cutoff value $\epsilon$, keep $\theta_p$ as a successful trial. 
\item Repeat steps 1---4 until sufficient number of trials have been successful.
\end{enumerate}

The distribution of parameters of the retained $\theta_p$ sample from successful trials approximates the posterior distributions of those parameters in the Bayesian framework, given $\epsilon$ is sufficiently small and the number of trials is sufficiently large. In practice, I set $\epsilon$ dynamically by sampling many times and saving each sampled $\theta_p$, then determining a value for $\epsilon$ which will retain less than some fraction of the total $\theta_p$ sample (in this work, this acceptance rate was set between 0.1\%---0.01\%, depending on the number of dimensions of the parameter space under consideration).

The similarity metric $\chi$ is usually defined by some distance on a set of summary statistics of the properties of $A$ and $B$, but implementations vary on a case-by-case basis. The use of summary statistics to define a metric of similarity is computationally advantageous for large datasets where other methods might be computationally prohibitive. If the summary statistics adopted are sufficient for the underlying distributions, then there is no information lost by using them. In this work, however, the data set is very small, and in the absence of a clear choice of sufficient summary statistics I instead utilize the two-sample Anderson-Darling test to determine a similarity metric (the two-sample AD-test $A_2$ statistic). For multiple dimensions, the adopted metric $\chi$ is the sum of all one-dimensional, two-sample $A_2$ statistics over all dimensions, divided by the number of dimensions tested. Qualitatively, this $\chi$ is small when all distributions are similar to each other, and large when one or more of the distributions are discrepant. In this implementation, the critical values of $A_2$ or $\chi$ and sample-size effects are accounted for as the ABCr analysis effectively bootstraps these out. The adoption of a metric based on a collection of 1D 2-sample AD-tests is motivated by its use by the CFEPS survey (eg., Petit et al. 2011), but by utilizing the ABCr framework instead of simply adopting the lowest $p$-value inferred from each independent AD-test over all dimensions, this work reduces the risk of erroneously identifying a low $p$-value in any given application of the metric.

\section{ The Orbit Distributions of the Jupiter Trojans }

The only other large population of Trojan objects in the solar system belongs to Jupiter. While evidently less populous than Neptune's Trojan swarms (Sheppard \& Trujillo 2006), the Jupiter Trojans are much better sampled, and are nearly complete to $H \lesssim 11$. To motivate the functional forms of inclination, libration amplitude, and eccentricity distributions I will adopt for the Neptune Trojans, I consider those of the Jupiter Trojans as templates.

\begin{figure*}[h]
\centering
\includegraphics[width=0.75\textwidth]{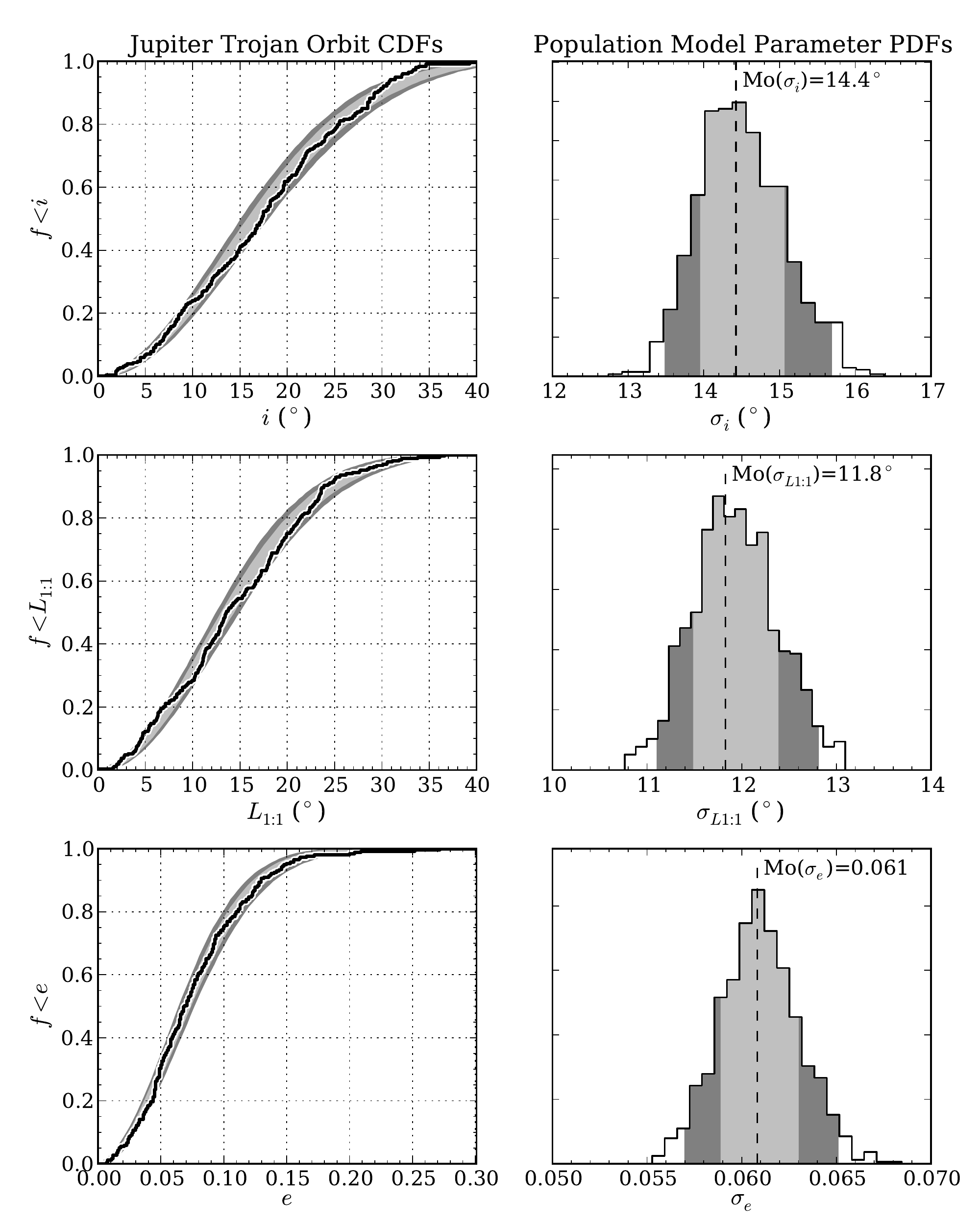}
\caption{Orbit distributions of a nearly complete, moderately bright sample of Jupiter Trojans ($H\geq11$), drawn from the Minor Planet Center (oscillating inclination and eccentricity) and AstDys-2 and the PETrA Project (libration amplitude). Inclination modeled by a truncated Brown's distribution, and both libration amplitude and eccentricity are modeled by truncated Rayleigh distributions. Light and dark gray model distributions illustrate model range covered 1- and 2-$\sigma$ intervals of accepted parameters.}
\label{JTPDF}
\end{figure*}

To date, no population-wide orbit distribution models have been computed for the Jupiter Trojans. Here I consider only the relatively complete $H \lesssim 11$ sample, and model the inclination, libration amplitude, and eccentricity distributions. 

I find that the inclination distribution is very well-modeled by a truncated Brown's distribution (Brown 2001):

\begin{equation}\label{id}
p ( i ) \propto  \begin{cases} \sin(i) e^{ -\frac{1}{2} \left( i / \sigma_i \right) ^2 }  di, & i < i_c \\ 
0, & i \geq  i_c, \end{cases}
\end{equation}

\noindent while both the libration amplitudes are well-modeled by truncated Rayleigh distributions defined by widths $\sigma_{L11}$ and $\sigma_e$ and truncations $L_{11c}$ and $e_c$. For the Jupiter Trojans I hold the truncations of each distribution fixed at the maximum observed value of each parameter; $i_c = 43.5^\circ$, $L_{11c}=36.645^\circ$, and $e_c = 0.272$. In the later Neptune Trojan analysis, these thresholds are defined by dynamical constraints. 

For consistency, I perform parameter estimation in the same ABCr framework that I later apply to the Neptune Trojans, though I only consider one distribution at a time. In the ABCr analysis, the parameters of interest ($\sigma_i$, $\sigma_{L11}$, and $\sigma_e$) are sampled uniformly in their distributions' means. That is to say, for each trial, $\langle i \rangle$ is selected uniformly, and then mapped to $\sigma_i$ given that the model inclination distribution is a Brown's distribution. Because the nominal functional form for the inclination, libration amplitude, and eccentricity distributions are truncated at some upper limit, their means are not trivial to compute. For example, the mean of a non-truncated Rayleigh distribution is simply given by $\sqrt{\pi/2}\sigma$, but for a truncated Rayleigh the mean is given by 

\begin{equation}
\langle x \rangle = \frac{ \sigma \sqrt{\frac{\pi}{2}}\mbox{erf}\left( x_t / \sqrt{2}\sigma \right) -x_t e^{-\frac{1}{2}(x_t/\sigma)^2}}{ 1 - e^{-\frac{1}{2}(x_t/\sigma)^2} },
\end{equation}

\noindent where $x_t$ is the truncation value (see Appendix \ref{RayMeanAppendix} for a derivation and more details). The functional form for a truncated Brown's distribution is similar (see Appendix \ref{BrownsMeanAppendix} for a derivation and more details). 

Proposing in the distribution mean has several advantages; chiefly, it translates an infinite prior volume ($ 0 \leq \sigma < \infty$) to a finite one, as the distribution mean is limited to a maximum value by the truncation of the distribution. For a truncated Rayleigh distribution, the mean asymptotically approaches $\langle x \rangle \rightarrow 2 x_t / 3$ as $\sigma \rightarrow \infty$. A truncated Brown's distribution has similar asymptotic behavior (see Appendices \ref{RayMeanAppendix} \& \ref{BrownsMeanAppendix} for details).

For the Jupiter Trojans, the uniform-mean prior is not particularly important because the sample is so large and well-defined. However, when I later consider the more poorly-sampled Neptune Trojans, the utility of this prior becomes much more apparent.

For each distribution, $10^6$ ABCr trials were run, and the 1,000 best trials were retained to generate the posterior PDF for each parameter. Figure \ref{JTPDF} illustrates the $H\leq11$ sample of observed osculating inclination, libration amplitude, and osculating eccentricity, compared to the accepted model distributions, and also illustrates the posterior PDFs for $\sigma_i$, $\sigma_{L11}$, and $\sigma_e$ defined by the accepted 1,000 ABCr trials. The observed Jupiter Trojan sample is very well modeled by the adopted functional forms for the orbit distribution; a KS-test indicates peak $p$-values of  92\%, 70\%, and 98\% for the inclinations, libration amplitudes, and eccentricities, respectively; indicating that there is no statistically significant evidence for discrepancy between the model distributions and the observed sample and qualitatively indicating good agreement between the two.

\section{Technique for debiasing the Neptune Trojans}\label{DebiasTech}

\subsection{Survey coverage}

Because of the lack of consistent information regarding their discovery surveys, a Canada-France Ecliptic Plane Survey (CFEPS)-style survey-simulator approach (Jones et al. 2006) is not feasible for the Neptune Trojans. Instead, I adopt a set of basis functions informed by typical trans-Neptunian object survey strategies which represent the coverage of the sky by all surveys which \emph{did discover} or \emph{plausibly could have discovered} Neptune Trojans. The parameters of these ``coverage functions'' are allowed to vary, and the discovery circumstances of each known Neptune Trojan are used to quantify how well a given choice of parameters reproduces them, given a set of assumed orbit distribution parameters. Finally, the properties of these coverage functions are marginalized over, and conservative orbit distribution posterior probability density functions are determined.

To model the {\it a priori} unknown survey properties, I assume that the probability of detecting a given Trojan is essentially independent of its orbital phase save for effects of areal coverage in ecliptic latitude $\beta$ and longitude $\lambda$. I fold all the effects of survey depth (which couples with the luminosity function) and areal coverage into two nuisance ``coverage functions'' $C(\beta)$ and $C(\lambda)$. These represent the probability of any survey detecting a Neptune Trojan given the Trojan's ecliptic latitude and longitude at the time of discovery. Nominally, these coverage functions will (1) only depend on ecliptic latitude and longitude, (2) likely be monotonic away from the centers of the Trojan clouds, and (3) have a functional range bracketed by the interval [0,1] and a domain over the entire sky.

In addition to modeling survey coverage in ($\lambda, \beta$), I also include the possibility of a direct bias against high-inclination objects in the MPC database. The CFEPS survey found moderately higher inclination distribution widths $\sigma_{i,32}$ for the 3:2 population than previous studies; because CFEPS carefully characterized their tracking biases in a self-consistent way, they suggested that earlier samples of objects with sufficiently accurate orbits to be confidently characterized as members of the 3:2 population may have had an unacknowledged bias against high inclinations. That such biases might exist is not surprising, and the possible sources of such biases were explored in Jones et al. (2006). Because the surveys that contributed to this inclination-biased sample were also sensitive to Neptune Trojans, it is likely that a similar inclination bias also exists in the observed sample of Neptune Trojans.

\subsubsection{Coverage function priors from Plutino population}

\begin{figure}[t]
\centering
\includegraphics[width=0.5\textwidth]{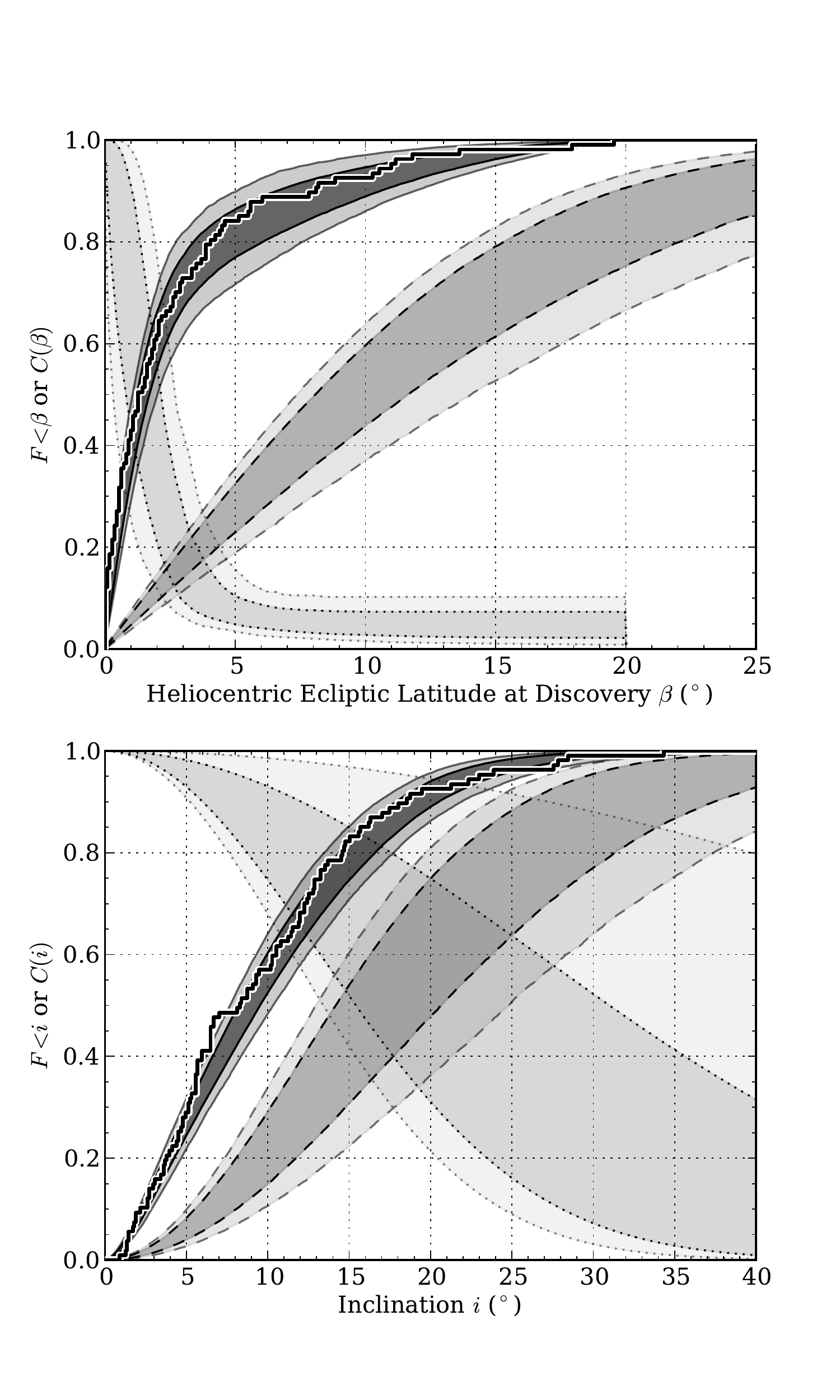}
\caption{Observed (solid histogram) 3:2 population Ecliptic latitude distribution (top panel) and inclination distribution (bottom panel), compared to an intrinsic distribution modeled by a Brown's inclination distribution (dashed regions) and biased by an Ecliptic latitude at discovery coverage function given by Eqn. \ref{LatCov} (top, dotted regions) and a Gaussian inclination coverage function (bottom, dotted regions), with parameters selected by procedure in text. The resulting biased model distributions (solid regions) excellently match the observed distribution, indicating that the general form of Eqn. \ref{LatCov} is a good model for the shape of the amalgamated latitude coverage functions of all previous surveys. 1- and 2-$\sigma$ ranges illustrated by dark and light regions, respectively.}
\label{plut_model}
\end{figure}

\begin{figure}[t]
\centering
\includegraphics[width=0.5\textwidth]{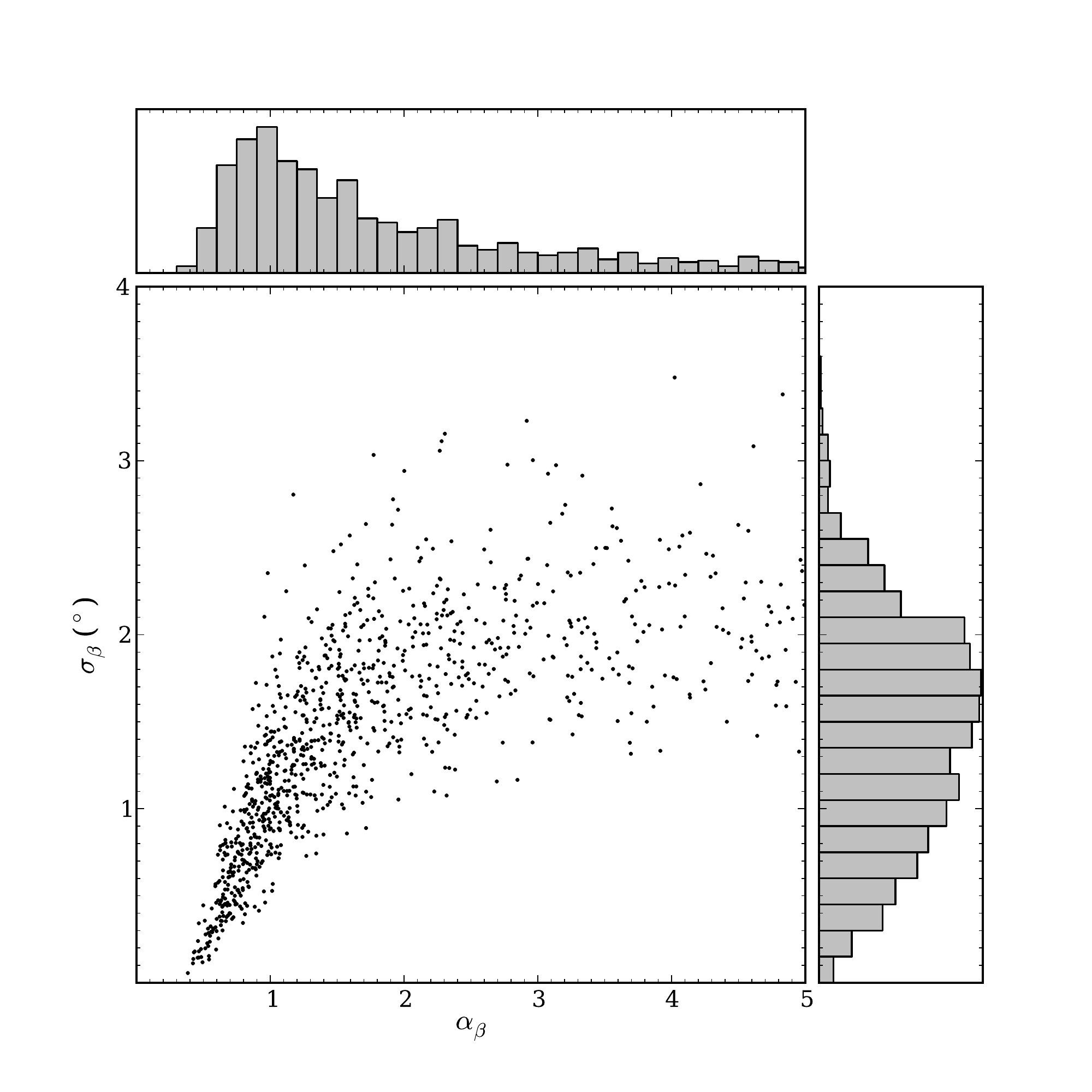}
\caption{ Retained sample of 3:2 population coverage function parameters $\sigma_\beta$ and $\alpha_\beta$ after ABCr analysis, and marginalized posterior PDFs for each parameter. }
\label{plut_modelAS}
\end{figure}

\begin{figure}[t]
\centering
\includegraphics[width=0.5\textwidth]{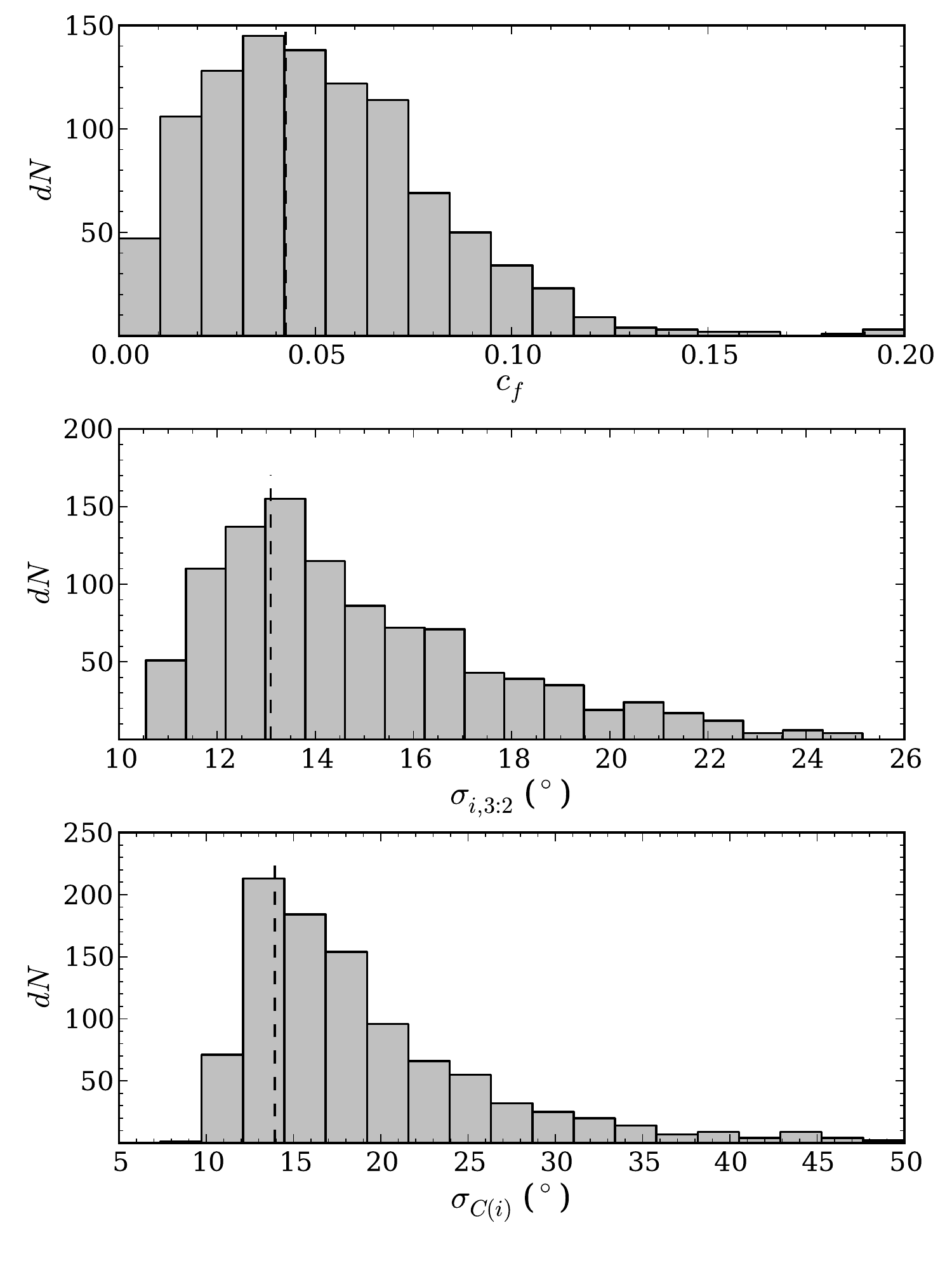}
\caption{ Marginalized posterior PDFs for other parameters in the 3:2 population ABCr analysis. Top panel: detection probability floor $c_f$. Middle panel: inclination distribution width $\sigma_{i,3:2}$. Bottom panel: inclination coverage function width $\sigma_{C(i)}$.  Vertical dashed lines indicate approximate distribution modes ($c_f=0.042$, $\sigma_{i,3:2}=13.1^\circ$, and $\sigma_{C(i)}=13.9^\circ$), estimated using a gaussian kernel density estimator. }
\label{plut_modelOTHERS}
\end{figure}


In order to determine the shape of the ecliptic latitude coverage function, I use the well-characterized 3:2 resonant population as a tracer. It is likely that similar ecliptic latitude and inclination biases apply to the 3:2 population as to the Neptune Trojan population; most surveys that discovered 3:2 objects were capable of discovering Neptune Trojans as well, as both populations share broad inclination distributions (and therefore both occupy a broad swath of ecliptic latitudes), similar on-sky rates of motion, and probably share similar luminosity functions.

The intrinsic properties of the 3:2 population have been measured by the carefully calibrated CFEPS survey (Petit et al. 2011, Gladman et al. 2012). By considering the \emph{observed} sample of (non-Kozai) 3:2 population members (provided by Buie, private communication), and adopting the intrinsic inclination distribution measured by CFEPS, it is possible to derive the longitudinally-averaged ecliptic coverage function that applied to the amalgam of all surveys that reported 3:2 discoveries to the Minor Planet center. 

I begin with an ansatz of an appropriate functional form for the ecliptic latitude coverage function $C(\beta)$:

\begin{equation}\label{LatCov}
C(\beta) = \begin{cases} 
\left(2^{- (\beta / \sigma_{\beta})^\alpha } + c_f \right)/\left(1 + c_f\right) & \beta \leq \beta_{3:2\mbox{max}} \\
0 & \beta > \beta_{3:2\mbox{max}}
\end{cases}
\end{equation}

This form has substantial flexibility provided by three free parameters, but the 3:2 observations constrain them to be strongly correlated. In general, its behavior for small $\beta$ is driven by index $\alpha$ and half-width $\sigma_{\beta}$, while for large $\beta$ the function flattens to a constant probability floor set by $c_f/(1+c_f)$. The function is truncated above $\beta \leq \beta_{32\mbox{max}}$, where $\beta_{32\mbox{max}}$ is the highest ecliptic latitude (at discovery) of any observed 3:2 object, since beyond this point the population provides little constraint on the properties of the coverage function.

I then consider the functional form of a possible inclination bias. In this case, there is useful information in literature to inform our functional ansatz: namely, that CFEPS and other surveys all found that the \emph{form} of the 3:2 population's inclination distribution was well-modeled by a Brown's distribution --- $p(i) \propto \sin(i) \exp(-i^2 / (2 \sigma_i^2) )$ --- but they found significantly different widths. In that case, I determine that the plausible form of the inclination bias is a Gaussian:

\begin{equation}
\sin(i) e^{-\frac{1}{2}(i/\sigma_{i1})^2} = C(i)\sin(i) e^{-\frac{1}{2}(i/\sigma_{i2})^2},
\end{equation}

\noindent so

\begin{equation}
C(i) = e^{ -\frac{i^2}{2} (\sigma_{i1}^{-2} - \sigma_{i2}^{-2} ) } = e^{-\frac{1}{2}(i/\sigma_{C(i)})^2},
\end{equation}

\noindent where $\sigma_{C(i)}^2 = 1 / (\sigma_{i1}^{-2} - \sigma_{i2}^{-2} )$. To map the best-fit CFEPS inclination width $\sigma_{i2} = 16^\circ$ to the best-fit DES inclination width $\sigma_{i2}=11^\circ$ (Gulbis et al. 2010), a Gaussian inclination coverage function requires a width of $\sigma_{C(i)} = \sqrt{ 1 / (16^{\circ -2} - 11^{\circ -2} ) } \simeq 15^\circ$. Gladman et al. (2012) notes that of 24 3:2 objects detected by CFEPS, four had inclinations over $21^\circ$, while no DES 3:2 objects had inclinations so high; adopting a Gaussian inclination coverage function width $\sigma_{C(i)}=15^\circ$, the probability of the observed sample retaining an otherwise observable object on an $i=21^\circ$ orbit is less than 40\%. At $i=34.4^\circ$ (the highest 3:2 inclination observed by CFEPS), this drops to roughly 7\%. 

The goal of the following analysis is to utilize the 3:2 population as a ``backlight'' for the Neptune Trojans, illuminating the biases that likely apply to both populations. Specifically, I do this by using the 3:2 population to inform the functional form of the coverage functions $C(\beta)$ and $C(i)$, and to derive priors for the parameters of these coverage functions. To do this, I used the sample of all objects in the MPC which are confidently identified as non-Kozai 3:2 objects by the DES classification scheme (Marc Buie, private communication). The observables I model are their inclination $i_O$ and their ecliptic latitude at discovery $\beta_O$. I model their inclination distribution as a Brown's distribution with width $\sigma_{i,32}$, and adopt the functional forms for $C(\beta)$ and $C(i)$ outlined above, and use ABCr to back out the PDFs of $\sigma_{i,32}$, $\sigma_{\beta}$, $\alpha$, $c_f$, and $\sigma_{C(i)}$. Given a sample of synthetic inclinations $i_{M}$ drawn from a model $M(\theta)$, synthetic ecliptic latitudes are generated by $\beta_{M} = \sin^{-1}( \sin(U) \sin(i_{M}) )$, where $U$ is a uniform random variate over $[0, 2\pi)$.

To define the prior on $\sigma_{i,32}$, I adopt the inclination distribution results for the Plutino population from the CFEPS analysis in Gladman et al. (2012); I approximate the posterior probability distribution for the Brown's distribution $\sigma_{i,32}$ as a triangular PDF with minimum value $10.55^\circ$, modal value $16^\circ$, and maximum value $25.96^\circ$. This triangular pdf reproduces the quoted mode from Gladman et al. (2012), 2.5\% of its integrated likelihood below their quoted 95\% lower confidence limit and 2.5\% of its integrated likelihood above their 95\% upper confidence limit. 

The parameter $\sigma_{\beta}$ is drawn from a uniform prior over the 5th and 95th percentiles of $\beta_O$ ($0.035 \leq \sigma_{\beta} \leq 10.90$). The parameter $\alpha$ is drawn from a uniform prior over $[0.2,5.0]$. The parameter $c_f$ was drawn from a uniform prior over $[0.0, 0.2]$. The parameter $\sigma_{C(i)}$ was defined as $\max(i_O) / \sqrt{ -\ln(f) }$, where $f$ was drawn from a uniform prior over $(0,1)$ --- this effectively defines the prior on $\sigma_{C(i)}$ such that the value of $C(\max(i_O))$ is drawn from a uniform prior.

Given these priors, the model for generating $i_M$ and $\beta_M$, and the observed samples of $\i_O$ and $\beta_O$, I perform $10^7$ ABCr trials with $\chi$ defined as the mean of $A_2(i_O, i_M)$ and $A_2(\beta_O, \beta_M)$, with equal sample sizes for the observed and model samples. The 1,000 trials which produced the smallest $\chi$ are reserved, and the distribution of $\sigma_{i,3}$, $\sigma_{\beta}$, $\alpha$, $c_f$, and $\sigma_{C(i)}$ which defined these trials defines the PDFs of each parameter. 

Given the simple population model, the proposed functional forms for $C(\beta)$ and $C(i)$ work remarkably well. Figure \ref{plut_model} illustrates the quality with which the model inclinations and ecliptic latitudes at discovery reproduce the observed distributions. Figures \ref{plut_modelAS} and \ref{plut_modelOTHERS} illustrate the marginalized PDFs of each parameter for the Plutino coverage function model that produced the excellent match illustrated in Figure \ref{plut_model}. These PDFs (except for that of $\sigma_{i,33}$) will become the priors for the coverage functions applied to the model Neptune Trojans in the next section.

\subsection{Longitude coverage function}

Using a similar procedure to determine the form of the longitude coverage function is less informative, as there may be a bias present for targeting the center of the Trojan clouds that is \emph{not} present in the discovered and reported 3:2 population.

Since I am considering two Trojan clouds, but am making the implicit assumption that both clouds have identical properties (aside from some relative occupation normalization, which does not affect this model), the observable derived from longitude I consider is the absolute value of the separation from the mean Trojan centers (defined for zero libration amplitude), or

\begin{equation}\label{Lprime}
\lambda' = | |\lambda - \lambda_N| - 60^\circ |
\end{equation}

I adopt a rollover function, normalized such that $C(0^\circ) = 1.0 $:

\begin{equation}\label{LonCov}
C(\lambda') = (1 - \tanh\left( W (\lambda' - \lambda'_c) \right))/(1 + \tanh\left( W \lambda'_c \right)).
\end{equation}

This functional form also has two free parameters (a turnover width $W$ and a turnover location $\lambda'_c$). To consider coverage ranging from effectively uniform in longitude to sharply biased toward the center of the Trojan swarms, I adopt a uniform distribution as the prior on $W$, allowing it to range from $0/^\circ$ (uniform coverage) to $+1/^\circ$ (strongly favoring detection toward the core of the Trojan clouds). $\lambda'_c$ is drawn from a uniform prior over $[0^\circ,10^\circ]$, roughly twice the range of all observed Neptune Trojan $\lambda'$.

\subsection{Neptune Trojan Orbit distribution forms}

The functional forms adopted for the inclination and libration amplitude distributions are motivated by the distributions of the Jupiter Trojans and other trans-Neptunian populations. The intrinsic inclination $i$ distribution is taken to be a truncated Brown's distribution (Brown 2001),

\begin{equation}\label{id}
p ( i ) \propto  \begin{cases} \sin(i) \exp( -\frac{1}{2} \left( i / \sigma_i \right) ^2 )  di, & i < i_t \\ 
0, & i \geq  i_t .\end{cases}
\end{equation}

Similarly, the intrinsic libration amplitude distribution is taken to be a truncated Rayleigh distribution, as was adopted for the distribution of libration amplitudes for the Jupiter Trojans),

\begin{equation}\label{ad}
p ( L_{11} ) \propto  \begin{cases} L_{11} \exp( -\frac{1}{2} \left( L_{11} / \sigma_{L 11} \right) ^2 )  dL_{11}, & L_{11} < L_{11t} \\ 
0, & L_{11} \geq  L_{11t} . \end{cases}
\end{equation}

Finally, the intrinsic eccentricity distribution is also taken to be a truncated Rayleigh distribution, as was adopted for the eccentricity distribution of Jupiter Trojans.

\begin{equation}\label{ad}
p ( e ) \propto  \begin{cases} e \exp( -\frac{1}{2} \left( e / \sigma_a \right) ^2 )  de, & e < e_t \\ 
0, & e \geq  e_t . \end{cases}
\end{equation}

These distributions are clipped by regions of dynamical instability. For the inclination distribution, there are no stable orbits above $i \geq60^\circ$ (Zhou, Dvorak \& Sun 2009), and in the following analysis the inclination threshold is set at $i_t = 60^\circ$. For the libration amplitude distribution, the truncation is placed at $L_{11t} = 35^\circ$ as larger libration amplitudes are unstable (Nesvorn\'{y} \& Dones, 2002). Eccentricities are conservatively limited to values below $e_t = 0.12$ (Zhou, Dvorak \& Sun 2010). Besides the physically-motivated fixed truncation locations, all of these these distributions are defined with one free parameter each.

\subsection{Pericenter bias}

For a variety of reasons, the eccentricity distribution of the Neptune Trojans is not likely to be strongly biased by observational effects. The eccentricity of Neptune Trojans are limited by stability constraints to relatively low values, $e \lesssim 0.12$; additionally, by definition, all Neptune Trojans have nearly identical semi-major axes. These two facts conspire to produce a relatively weak pericenter bias. I demonstrate this by considering the relative detection rate of the eccentricity extrema of the population using a simple Monte Carlo test: draw two large samples of $H$-magnitudes from a power-law luminosity function with slope $\alpha$ and maximum value $H_0$. One sample I preserve, and take it to represent the apparent magnitudes of objects on circular orbits $m_0$. To the other sample I add a quantity $dm = 10 \log_{10}(r)$ where $r$ is drawn from a sample of Keplerian orbits with unit semi-major axis,uniformly distributed Mean Anomaly, and non-zero, finite eccentricity --- this sample represents the apparent magnitudes of objects on eccentric orbits $m_e$. I then define a detection probability function of similar form to real survey sensitivities, 

\begin{equation}
p(m) = \frac{1}{2} [ 1-\tanh ( (m - m_{50})/w ) ],
\end{equation}

\noindent where $0.3 \lesssim w \lesssim 1.0$ for typical ground based surveys. Amalgams of multiple surves tend to have broader turnovers still. I select a value for $w$ and set $m_{50} = H_0 - 3w$. The potential for bias is estimated by determining the ratio of the detection rate of objects on eccentric orbits to the detection rate of objects on circular orbits.

The results of this simple analysis are illustrated in Fig. \ref{ecc_bias}; for reasonable power-law slopes, the expected preference for eccentric orbits over circular ones never exceeds 20\% over unity --- and for the most eccentric \emph{known} Neptune Trojan, the preference never exceeds 10\%. 

In addition, the Neptune Trojan luminosity function has been observed to break to very shallow slope at $m_R \simeq 23.5$ (Sheppard \& Trujillo 2010a). If a survey reaches fainter than this break, it effectively ``punches through'' the entire Trojan population and is complete over all heliocentric distances occupied by them. The majority of discovered Neptune Trojans were found in surveys that reached substantially fainter than this (six from Sheppard \& Trujillo 2006 \& 2010b, one from Parker et al. 2013). For the purposes of the present analysis, then, I proceed assuming that the eccentricity bias in the detected sample is weak and can be neglected.

\begin{figure}[t]\label{ecc_bias}
\centering
\includegraphics[width=0.5\textwidth]{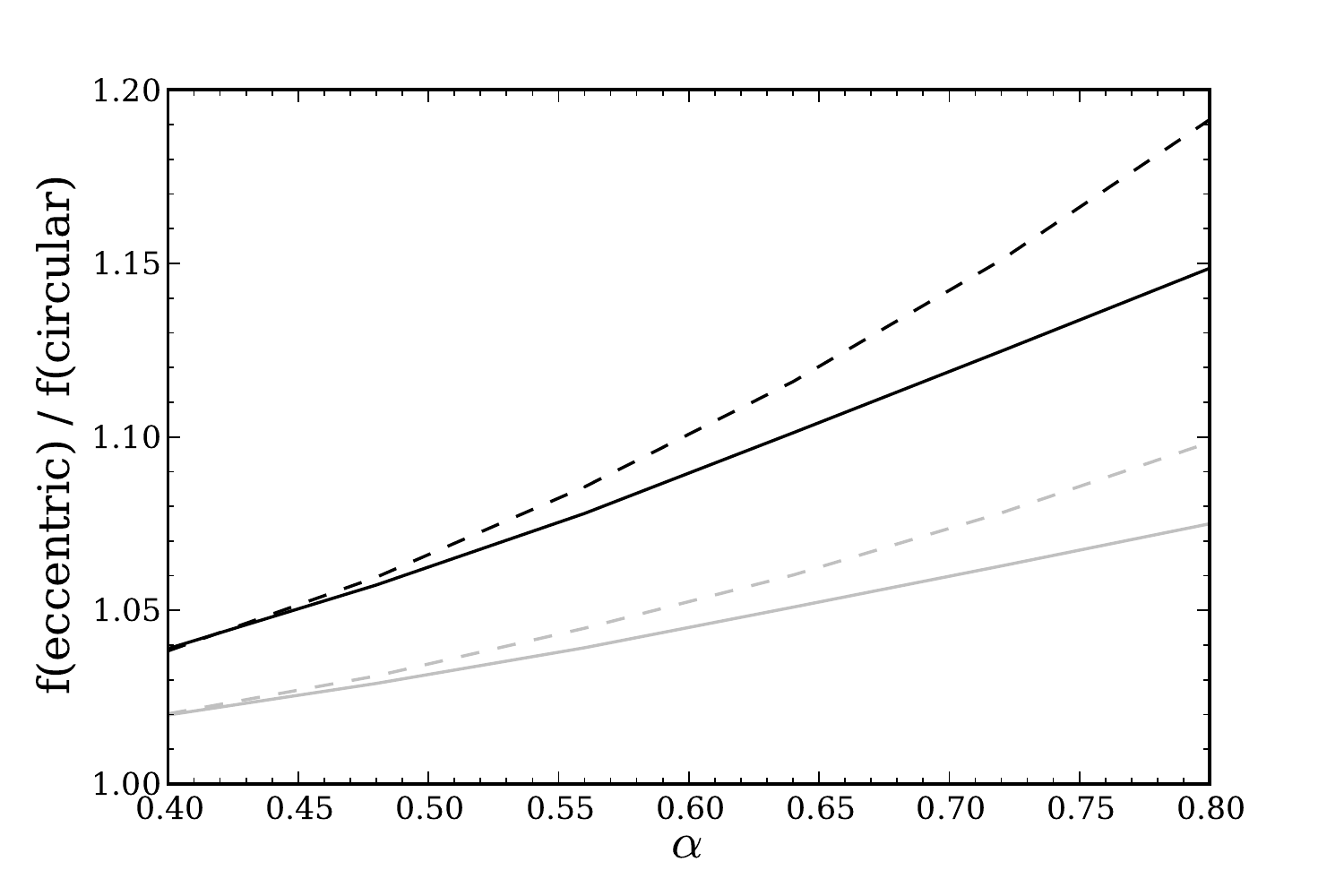}
\caption{Potential detection rate bias for eccentric orbits as function of luminosity function slope $\alpha$: black lines illustrate ratio of detection rates for $e=0.12$ orbits (roughly the maximum stable Neptune Trojan eccentricity) over circular orbits, while gray lines illustrate ratio of detection rates for $e=0.086$ orbits (the maximum observed eccentricity) over circular orbits. Dashed line corresponds to a survey flux-sensitivity rollover width of $w=0.3$ magnitudes, and the solid line to a broader $w=1$ magnitude rollover width. }
\label{ecc_bias}
\end{figure}

\subsection{Priors}

As with the Jupiter Trojans, for the main parameters of interest ($\sigma_i$, $\sigma_{L11}$, and $\sigma_e$) I propose uniformly in their distributions' means. This becomes particularly important for the Neptune Trojans because of the inversion of an infinite prior volume to a finite one, allowing the consideration of very large widths in a meaningful way and without imposing an artificial cutoff to the proposal volume.

The latitude and inclination coverage function parameters are drawn randomly from the retained sample of 1,000 Plutino model trials. This preserves all correlations present in the posteriors of these parameters when informed by the observed Plutino population. The longitude coverage function's inverse rollover width $W$ is drawn from a uniform prior over  [$0/^\circ$, $+1/^\circ$], while $\lambda'_c$ is drawn from a uniform prior over $[0^\circ,11^\circ]$, containing the range of all observed Neptune Trojan $\lambda'$.

\subsection{ Sample Generation }

Given the orbit distributions and priors described in the previous sections, synthetic samples of Neptune Trojans can be generated. The model I use to generate the synthetic properties and discovery circumstances for the Neptune Trojans is described in detail in Appendix B; in brief, it takes as input the orbit distribution model parameters $\sigma_i$, $\sigma_{L11}$, $\sigma_e$, $i_t$, $L_{11t}$, and $e_t$, and the coverage function forms and parameters, and returns a sample of synthetic ($i,L_{11},e,\beta,\lambda'$) that reflects both the model intrinsic distributions and the model biases simulated by the coverage functions.

\section{Neptune Trojan Orbit Distribution Results}

\begin{table*}[t]\label{ResultsTable}
\centering
\begin{tabular}{  c c c c c c }
\multicolumn{6}{c}{\bf Table 2: Neptune Trojan Orbit Distribution Parameter Confidence Intervals}\\
\hline
\hline
Parameter & Mode & Lower 68\% & Upper 68\% & Lower 95\% & Upper 95\% \\
\hline

$\sigma_L$  & 10$^\circ$   & 9$^\circ$  &  16$^\circ$   &  7$^\circ$ & 26$^\circ$\\
$\sigma_e$  & 0.044  & 0.039  &  0.070   &  0.033 & 0.125\\

\multicolumn{6}{c}{\bf With Inclination Bias }\\
$\sigma_i$  & 21$^\circ$   & 17$^\circ$  &  46$^\circ$   &  13$^\circ$ & 91$^\circ$\\

\multicolumn{6}{c}{\bf Without Inclination Bias }\\
$\sigma_i$  & 17$^\circ$  & 14$^\circ$  &  26$^\circ$   &  11$^\circ$ &  46$^\circ$ \\

\hline
\end{tabular}
\end{table*}

Parameter estimation was performed in the ABCr framework. For each case under consideration, $10^8$ ABCr trials were run, and the 10,000 best trials from each were retained to generate the marginalized posterior PDFs for each parameter under consideration. Two cases were considered: in case (1), I included the effect of the inclination coverage function informed by the Plutinos. In case (2), the inclination coverage function was ignored. This allows the effect of the putative inclination bias on the results to be quantified. 

\begin{figure}
\centering
\includegraphics[width=0.5\textwidth]{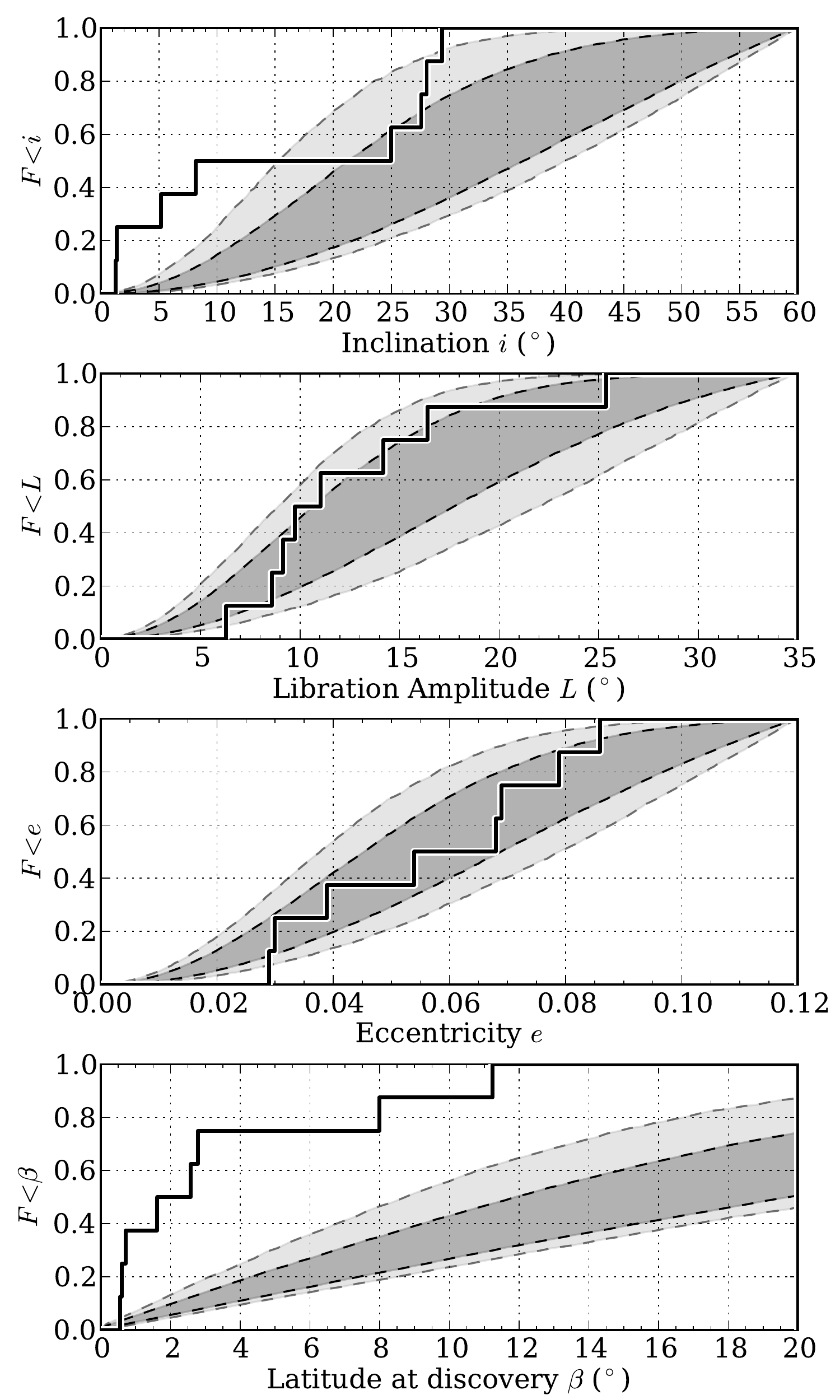}
\caption{Intrinsic (debiased) distributions for Neptune Trojans, drawn from accepted model trials with inclination bias allowed. Black histogram illustrates observed population (with biases). Dark and light gray ranges show 1- and 2-$\sigma$ range of accepted models. }
\label{IntWithBias}
\end{figure}

\begin{figure}
\centering
\includegraphics[width=0.5\textwidth]{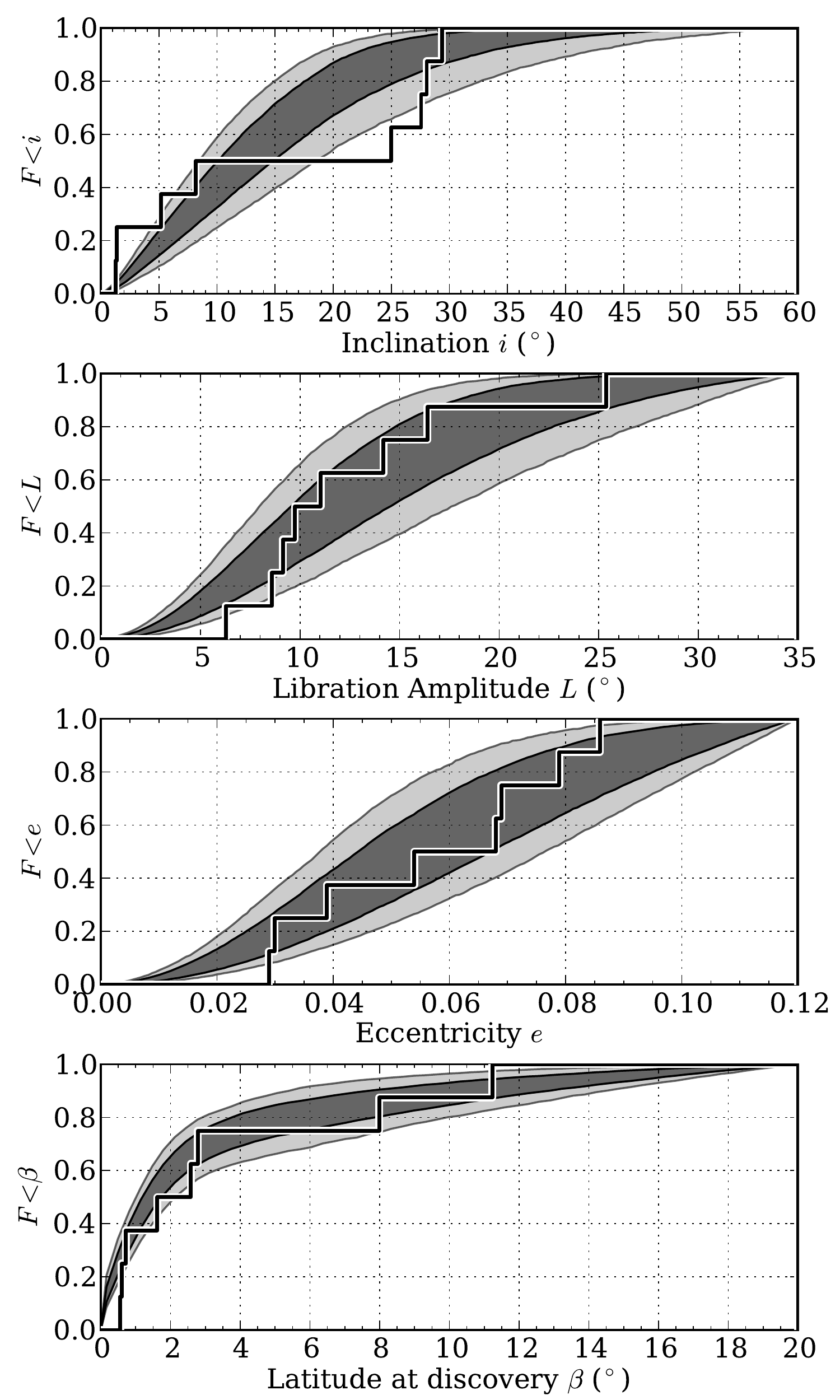}
\caption{Biased distributions for Neptune Trojans, drawn from accepted model trials with inclination bias allowed. Black histogram illustrates observed population (with biases). Dark and light gray ranges show 1- and 2-$\sigma$ range of accepted models. }
\label{PrimeWithBias}
\end{figure}

\begin{figure}[t]
\centering
\includegraphics[width=0.5\textwidth]{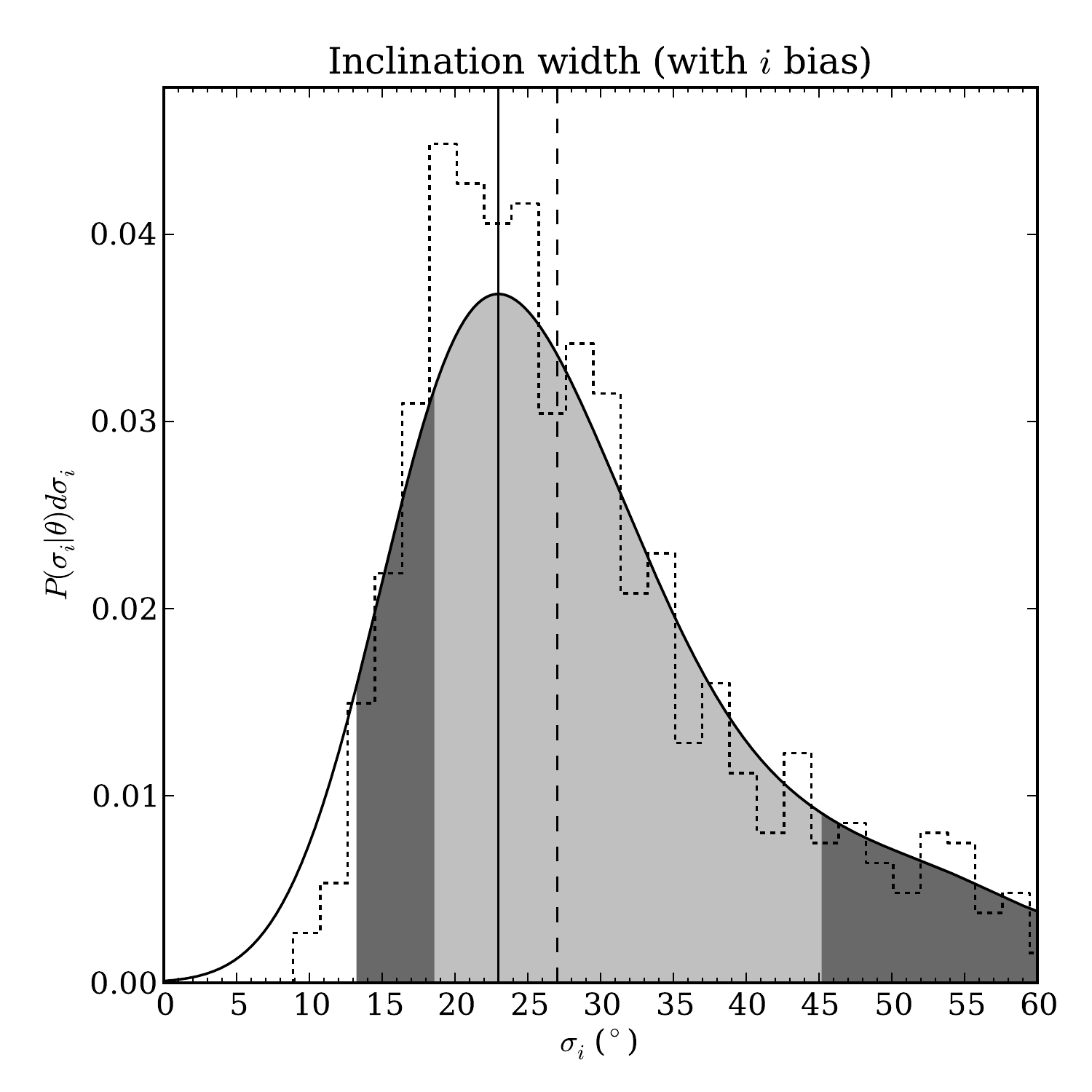}
\includegraphics[width=0.5\textwidth]{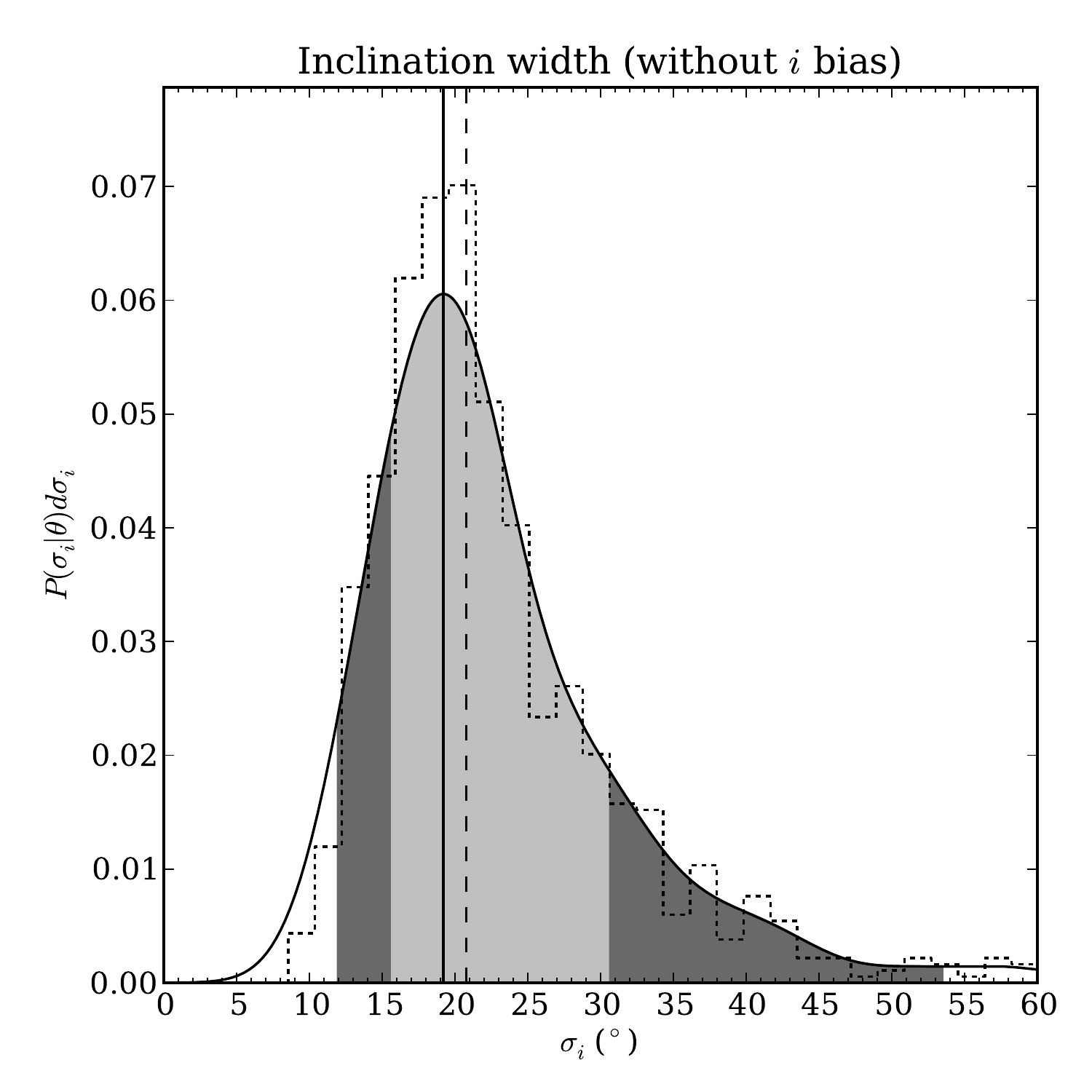}
\caption{Posterior distribution of Neptune Trojan inclination width parameter $\sigma_i$, top panel illustrating the distribution when allowing a gaussian inclination bias, and bottom panel when such a bias is not allowed for. Shaded regions illustrate 68\% and 95\% confidence intervals determined directly from distribution of accepted ABCr samples; solid vertical line marks mode estimated by a gaussian Kernel Density Estimator, dashed vertical line marks distribution median. }
\label{fig:IncPosteriors}
\end{figure}

\begin{figure}
\centering
\includegraphics[width=0.5\textwidth]{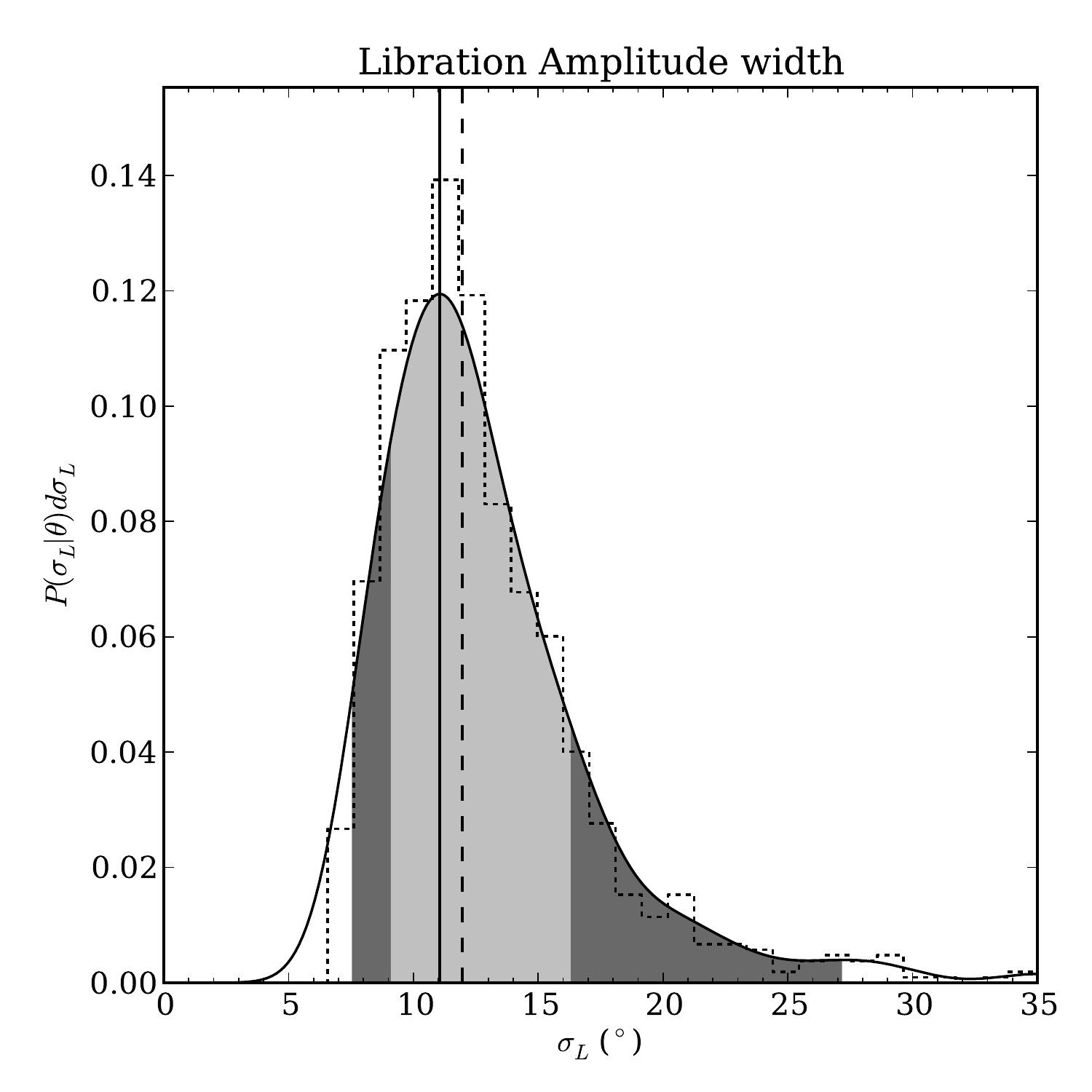}
\caption{Same as Figure \ref{fig:IncPosteriors}, but for libration amplitude width $\sigma_{L11}$. No distinction between models including or ignoring inclination bias, so only one PDF is illustrated.}
\label{fig:AmpPosteriors}
\end{figure}

\begin{figure}
\centering
\includegraphics[width=0.5\textwidth]{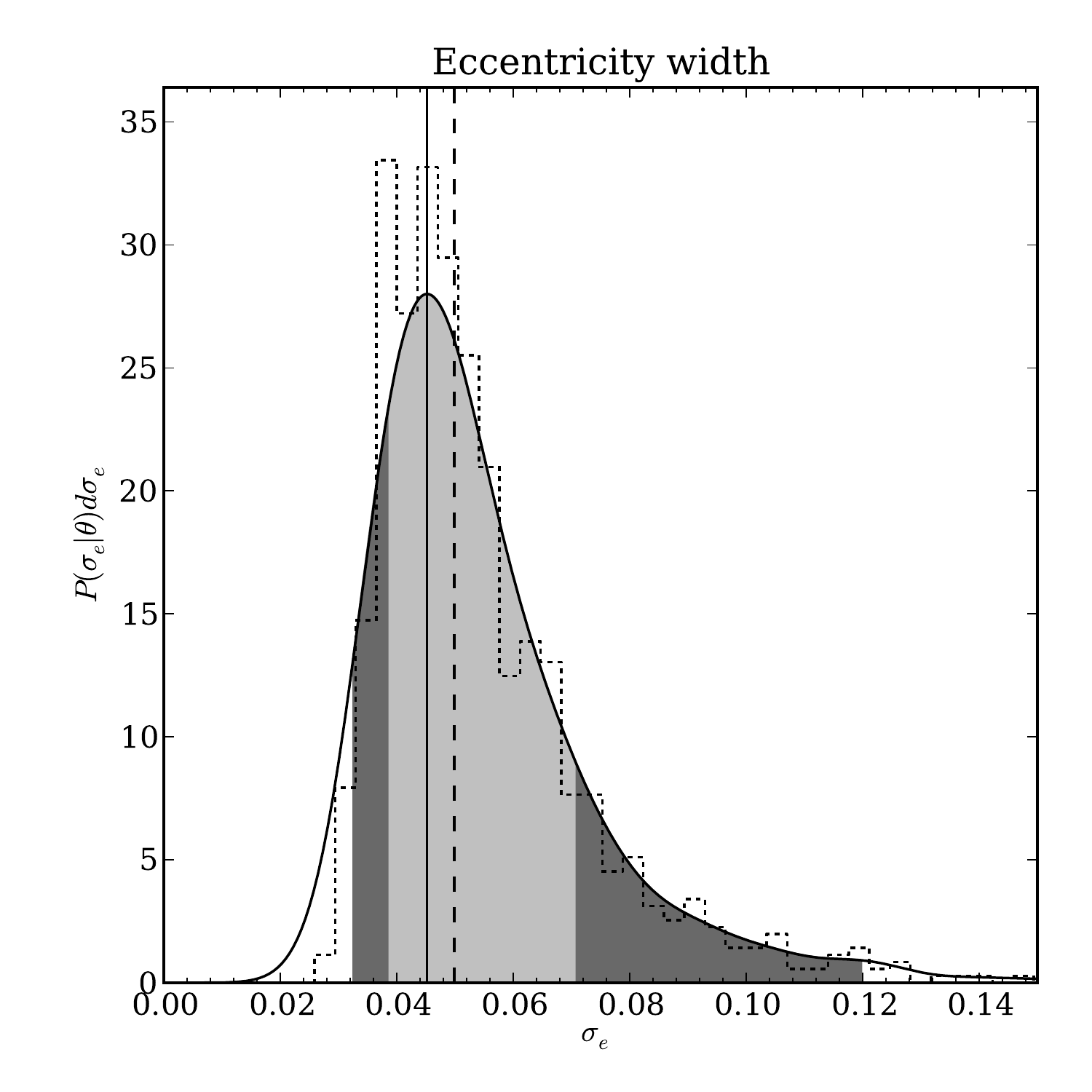}
\caption{Same as Figure \ref{fig:IncPosteriors}, but for eccentricity width $\sigma_{e}$. No distinction between models including or ignoring inclination bias, so only one PDF is illustrated.}
\label{fig:EccPosteriors}
\end{figure}

\subsection{Case 1: With inclination bias}

Figure 7 illustrates the allowed intrinsic distributions of Neptune Trojan orbit properties and observables, while Figure 8 illustrates the allowed apparent distributions of these same properties after the survey coverage functions are applied. The most drastic effect of these coverage functions is apparent in the distribution of latitude at discovery; while the intrinsic distribution is substantially different from the observed sample, the biased distribution matches the observed sample very well (much as it did for the 3:2 sample). 

With the inclination coverage function drawn from the Plutino priors, the width of the Neptune Trojan inclination distribution is allowed to climb to very large values ($\sim91^\circ$) within the 95\% confidence interval, effectively allowing solutions approaching a uniform $p(i) \propto \sin(i)$ distribution. The lower limit on the 95\% confidence interval is $\sigma_i \geq 13^\circ$, and the mode of the posterior PDF (illustrated in Figure 9) is $\sigma_i = 21^\circ$. This is admittedly a large range; but even so it is a formal definition of the uncertainty when adopting a Brown's distribution for the inclination distribution, and can be utilized for studying the origins of the population. 

While their overall distribution is reasonably well-reproduced by the models adopted here,  the observed Neptune Trojan inclinations appear somewhat bi-modal, perhaps suggesting structure more complex than the simple single Brown's distribution used. Preliminary tests did not indicate any statistical evidence for a more complex two-component Brown's distribution (similar to the superimposed hot and cold classical Kuiper Belt), but such a model cannot be ruled out with the current sample. 

The libration amplitude distribution is somewhat better constrained at its upper end; the maximum allowed value of $\sigma_{L 11}$ within the 95\% confidence interval is $26^\circ$, which is nine degrees shy of the truncation limit of the libration amplitudes, indicating a strong requirement for the presence of the gaussian component of the Rayleigh distribution. Widths as low as $\sigma_{L 11} = 7^\circ$ are allowed, and the peak of the distribution, $\sigma_{L 11} = 10^\circ$ is very similar to that of the Jupiter Trojans ($11.8^\circ$). The posterior PDF for the accepted libration amplitude distribution $\sigma_{L 11}$ parameter is illustrated in Figure 10.

The eccentricity distribution is a more moderate case; if a slightly lower eccentricity truncation were adopted (eg., 0.08, Nesvorny \& Vokrouhlicky 2009) the distribution would be consistent with $p(e)\propto e$. With the adopted truncation of $e\leq0.12$, the maximum distribution width allowed in the 95\% confidence interval is $\sigma_e \leq 0.125$. The lower limit is $\sigma_e \geq 0.033$, and the peak falls at $\sigma_e = 0.044$, similar again to the Jupiter Trojans ($\sigma_e = 0.061$). The posterior PDF for the accepted eccentricity distribution $\sigma_e$ parameter is illustrated in Figure 11.

\subsection{Case 2: Without inclination bias}

When the inclination coverage function is set to unity over all inclinations, and all other priors remain as before, the entire confidence interval for the inclination distribution width is shifted to lower widths. The upper limit of the 95\% confidence interval becomes better constrained, at $\sigma_i \leq 46^\circ$, significantly lower than the truncation inclination and indicating evidence for the presence of the gaussian component of the Brown's distribution. The lower limit drops by a few degrees to $\sigma_i \geq 11^\circ$, and the mode of the PDF (illustrated in the lower panel of Figure 9) decreases to $\sigma_i = 16^\circ$, similar to the nominal width of the Plutinos ($16^\circ$, Gladman et al. 2012) and the Jupiter Trojans ($14.4^\circ$).

All other distributions were unaffected, and retained identical posterior PDFs.

\section{Simulated Capture of Neptune Trojans}

\begin{table*}
\begin{tabular}{ cccccccccccc }
\multicolumn{12}{c}{\bf Table 3: Synthetic Captured Trojan Properties }\\
\hline
ID & $a_{N0}$ & $a_{U0}$ & $\sigma_{i0}$ & $e_{N0}$ & $\tau_a$ & Capture Eff. & $N$ Captured & $\sigma_{i}$ & $\sigma_{L11}$ & Peak $i$ prob. & Peak $L_{11}$ prob. \\
\hline

1 & 26 & 17.820 & 10$^\circ$ & 0.05 & 1$\times10^{6}$ &  3.20$\times10^{-4}$ & 8 & $14.38^{+1.97}_{-1.55}$ & $13.68^{+11.20}_{-1.04}$ & 0.335 & 0.935 \\
2 & 26 & 17.820 & 10$^\circ$ & 0.1 & 1$\times10^{6}$ &  2.52$\times10^{-3}$ & 63 & $9.26^{+0.54}_{-0.60}$ & $30.92^{+14.27}_{-3.27}$ & 0.544 & 0.972 \\
3 & 26 & 17.820 & 10$^\circ$ & 0.15 & 1$\times10^{6}$ &  1.74$\times10^{-3}$ & 87 & $11.94^{+0.48}_{-0.77}$ & $19.13^{+2.68}_{-1.19}$ & 0.745 & 0.158 \\
4 & 26 & 17.820 & 10$^\circ$ & 0.15 & 1$\times10^{6}$ &  8.80$\times10^{-4}$ & 22 & $9.67^{+1.73}_{-0.83}$ & $13.23^{+1.78}_{-1.29}$ & 0.984 & 0.425 \\
5 & 26 & 17.820 & 10$^\circ$ & 0.2 & 1$\times10^{6}$ &  6.00$\times10^{-4}$ & 15 & $9.85^{+1.79}_{-1.55}$ & $14.67^{+5.45}_{-1.39}$ & 0.658 & 0.841 \\
6 & 26 & 17.820 & 15$^\circ$ & 0.15 & 1$\times10^{6}$ &  1.00$\times10^{-3}$ & 25 & $16.16^{+1.43}_{-1.55}$ & $28.15^{+14.12}_{-5.15}$ & 0.739 & 0.837 \\
7 & 26 & 17.820 & 2$^\circ$ & 0.15 & 1$\times10^{6}$ &  3.32$\times10^{-3}$ & 83 & $2.47^{+0.12}_{-0.18}$ & $13.58^{+0.50}_{-0.59}$ & 0.861 & 0.144 \\
8 & 26 & 17.820 & 5$^\circ$ & 0.15 & 1$\times10^{6}$ &  2.52$\times10^{-3}$ & 63 & $4.37^{+0.42}_{-0.30}$ & $20.32^{+3.17}_{-1.39}$ & 0.926 & 0.461 \\
9 & 28 & 19.191 & 10$^\circ$ & 0.0086 & 1$\times10^{6}$ &  2.67$\times10^{-5}$ & 2 & --- & --- & --- & --- \\
10 & 28 & 19.191 & 10$^\circ$ & 0.05 & 1$\times10^{6}$ &  1.20$\times10^{-3}$ & 30 & $10.92^{+1.61}_{-0.89}$ & $15.27^{+3.72}_{-1.09}$ & 0.993 & 0.985 \\
11 & 28 & 19.191 & 10$^\circ$ & 0.10 & 1$\times10^{6}$ &  4.00$\times10^{-5}$ & 1 & --- & --- & --- & --- \\
12 & 28 & 19.191 & 10$^\circ$ & 0.15 & 1$\times10^{6}$ &  5.40$\times10^{-4}$ & 54 & $10.80^{+0.83}_{-0.60}$ & $11.10^{+1.24}_{-0.45}$ & 0.366 & 0.813 \\
13 & 28 & 19.191 & 10$^\circ$ & 0.15 & 1$\times10^{6}$ &  3.20$\times10^{-4}$ & 8 & $9.73^{+5.84}_{-0.54}$ & $8.28^{+1.54}_{-0.84}$ & 0.912 & 0.608 \\
14 & 28 & 18.531 & 10$^\circ$ & 0.1 & 1$\times10^{6}$ &  5.20$\times10^{-4}$ & 26 & $11.16^{+1.13}_{-1.07}$ & $12.74^{+2.03}_{-0.69}$ & 0.964 & 0.852 \\
15 & 28 & 18.531 & 10$^\circ$ & 0.2 & 1$\times10^{6}$ &  2.13$\times10^{-4}$ & 16 & $20.51^{+2.03}_{-2.14}$ & $16.26^{+8.62}_{-1.29}$ & 0.273 & 0.884 \\
16 & 28 & 17.870 & 10$^\circ$ & 0.15 & 1$\times10^{6}$ &  1.60$\times10^{-3}$ & 40 & $11.82^{+0.71}_{-0.71}$ & $14.97^{+4.26}_{-0.64}$ & 0.325 & 0.879 \\
17 & 28 & 19.191 & 15$^\circ$ & 0.15 & 1$\times10^{6}$ &  2.00$\times10^{-4}$ & 10 & $21.05^{+6.49}_{-1.91}$ & $10.01^{+1.78}_{-0.94}$ & 0.896 & 0.705 \\
18 & 28 & 19.191 & 2$^\circ$ & 0.15 & 1$\times10^{6}$ &  7.20$\times10^{-4}$ & 18 & $4.37^{+0.60}_{-0.83}$ & $9.42^{+1.34}_{-1.24}$ & 0.779 & 0.998 \\
19 & 28 & 19.191 & 5$^\circ$ & 0.15 & 1$\times10^{6}$ &  7.48$\times10^{-4}$ & 56 & $6.10^{+0.60}_{-0.30}$ & $12.44^{+0.59}_{-1.29}$ & 0.940 & 0.599 \\
20 & 28 & 18.531 & 8$^\circ$ & 0.2 & 1$\times10^{6}$ &  8.00$\times10^{-5}$ & 8 & $15.33^{+2.50}_{-1.85}$ & $19.18^{+21.50}_{-1.34}$ & 0.396 & 0.969 \\
21 & 28 & 19.191 & 10$^\circ$ & 0.15 & 1$\times10^{7}$ &  5.60$\times10^{-4}$ & 28 & $13.66^{+0.83}_{-1.01}$ & $12.84^{+1.54}_{-0.69}$ & 0.378 & 0.381 \\
22 & 28 & 19.191 & 2$^\circ$ & 0.15 & 1$\times10^{7}$ &  3.60$\times10^{-4}$ & 27 & $4.19^{+0.54}_{-0.36}$ & $14.97^{+2.87}_{-0.94}$ & 0.539 & 0.784 \\
23 & 28 & 19.191 & 5$^\circ$ & 0.15 & 1$\times10^{7}$ &  7.60$\times10^{-4}$ & 19 & $8.48^{+0.95}_{-0.95}$ & $12.69^{+2.38}_{-0.99}$ & 0.619 & 0.839 \\
24 & 28 & 19.191 & 10$^\circ$ & 0.15 & 5$\times10^{6}$ &  5.40$\times10^{-4}$ & 27 & $12.47^{+1.07}_{-0.83}$ & $17.10^{+4.66}_{-1.39}$ & 0.467 & 0.690 \\
25 & 28 & 19.191 & 10$^\circ$ & 0.15 & 1$\times10^{6}$ &  6.67$\times10^{-5}$ & 5 & $13.90^{+7.39}_{-1.31}$ & $15.76^{+18.38}_{-1.54}$ & 0.616 & 0.764 \\

\hline
\end{tabular}
\label{CaptureResults}

\vspace{30pt}

\end{table*}

In order to determine the implications of the measured properties of the Neptune Trojan orbit distribution, I performed an exploratory suite of Neptune Trojan capture simulations. These simulations were similar to those of Lykawka et al. (2009), except that I include an initial eccentric epoch for Neptune and consider a pre-excited disk. These simulations and their outcomes are detailed in the following section.

These simulations were performed with a version of \textit{mercury6} (Chambers 1999) that includes an artificial, user-defined force designed to drive semi-major axis migration and eccentricity damping (Wolff et al. 2012). The semi-major axis evolution followed an exponential form 

\begin{equation}
a(t) = a_i + (a_i - a_f)\exp(-t/\tau_a),
\end{equation}

\noindent where $a(t)$ is a given planet's semi-major axis at time $t$ in the integration, $a_i$ and $a_f$ represent the planet's initial and final semi-major axes, respectively, and $\tau_a$ is a characteristic timescale for migration.

Eccentricity damping was performed in a slightly more general way than has been implemented in previous works. I use an identical functional form as for semi-major axis migration,

\begin{equation}
e(t) = e_i + (e_i - e_f)\exp(-t/\tau_e),
\end{equation}

\noindent where $e(t)$ is a given planet's eccentricity at time $t$ in the integration, $e_i$ and $e_f$ represent the planet's initial and final eccentricities, respectively, and $\tau_a$ is a characteristic timescale for eccentricity damping. This allows me to damp eccentricity asymptotically to some final, finite value. Because I am interested in the long-term behavior and stability of captured Trojans, I wish to replicate the final architecture of the giant planets' orbits as closely as possible. As such, this ability to drive eccentricities to values near their current vaues (instead of to zero) is important.

The initial configuration of the giant planets was motivated by the Nice Model and by the recent work of Wolff et al. (2012) and Dawson \& Murray-Clay (2012), which argued that in order to preserve the cold classical Kuiper Belt while populating the hot classical Kuiper Belt, Neptune must have had a significant eccentricity ($e_N>0.15$) at the start of its final epoch of smooth outward migration. This initial eccentricity must have damped away relatively quickly, with a characteristic timescale less than a few times $10^5$ years. They argue that the characteristic timescale for semi-major axis must have been longer than the characteristic timescale for eccentricity damping.

In these simulations, Neptune is started with a moderate initial eccentricity. However, in order to avoid early scattering interactions with Uranus that would add an unwanted stochastic element to the outcome of these simulations, for many initial configurations Neptune's eccentricity is not set as high as dictated by the limits of Dawson \& Murray-Clay (2012) given Neptune's initial semi-major axis.

The disks that Neptune migrates into are composed of massless tracer particles. The disks are started with a Brown's distribution for inclination, a Rayleigh distribution for eccentricity, and a power-law semi-major axis distribution with index $-3/2$ (approximately producing a $r^{-5/2}$ surface density distribution) defined over the range of 22 AU $< a <$ 30 AU. The width of the eccentricity distribution is scaled from of the width of the inclination distribution, such that $\sigma_e = \sin(\sigma_i)$. If any object is generated with a pericenter lower than Saturn's initial semi-major axis, it is re-drawn. Initial inclination widths between $2^\circ$ and $15^\circ$ were explored. The smallest disk considered was populated with 25,000 particles, while the largest was populated with 100,000 particles.

Two initial semi-major axes were considered ($a_{N0}=$26 AU and 28 AU), and the initial semi-major axis of Uranus was set in each case to sweep through a desired range of Neptune:Uranus period ratios. Since Uranus and Neptune are not trapped in their 2:1 MMR today, it is unlikely that they ever passed through it. As such, due to their current proximity to the 2:1, the parameter volume from which the system could migrate through \emph{decreasing} period ratios is very small, and I only consider initial configurations where the Neptune:Uranus period ratio increased (or remained constant) with time. At one extreme, Uranus is placed as close to its current location as possible without causing early-time interactions with the initial state of Neptune. For the $a_{N0} = 28$ AU simulation, this places Uranus at its current semi-major axis. At the other extreme, Uranus was placed such that over the course of the simulation, the Neptune:Uranus period ratio did not evolve. For equal $\tau_a$, this is trivially achieved by setting $a_{U0} = a_{U1} a_{N0}/a_{N1}$. Several intermediate cases were explored as well.

All giant planets other than Neptune experienced no eccentricity migration; their initial eccentricities were set at values representative of their current mean values. For Neptune, regardless of initial eccentricity, the eccentricity damping timescale $\tau_e$ was set to $3 \times 10^5$ years.

The integrations were performed using \emph{mercury}'s hybrid integrator, and run for $5 \tau_a$; at this point, the disk particles were down-selected to only those with semi-major axes in proximity to Neptune's (semi-major axis within one Neptune Hill radius). The integrations were then continued with only the giant planets and the down-selected disk particles until 100 Myr total time had elapsed. For two small simulations, the integration was continued for 1 Gyr. After the prime integration was complete, a short 100,000 year integration was performed where each particle's state vectors were saved very frequently. From this 100,000 year integration, the properties of the remaining objects were determined; if they were librating around the L4 or L5 points of Neptune with libration amplitudes less than $35^\circ$ and had eccentricities less than 0.12, they were identified as stable Neptune Trojans. 

The number and properties of the objects identified as stable Neptune Trojans are compiled for each simulation and modeled with the same distributions used to model the observed sample in the preceding sections. Table 3 outlines the initial conditions of each simulation, the number of Trojans captured and the implied capture efficiency, and the properties of the inclination and libration amplitude distributions of the captured sample. Capture efficiencies are inferred from the structure of the primordial disk imposed; since Neptune may only sweep up Trojans from a small radial extent within the disk, these efficiencies could have strong spatial variance and simply using a different radial disk structure could produce substantially different capture efficiencies. These efficiencies ($10^{-5}-10^{-3}$), are often sufficiently high that Trojan populations captured through the processes modeled here could be substantially reduced without being problematically low compared to the extant population size (Nesvorn\'{y} \& Vokrouhlick\'{y} 2009).

Figure \ref{all_sim_results} illustrates the initial conditions and capture results of all 25 simulations. In general, it is clear that for most simulations, particles prefer to become captured onto Trojan orbits with inclinations similar to their primordial inclinations, and the mean primordial inclinations of particles that end up becoming Neptune Trojans generally closely match their mean inclinations as Neptune Trojans. This will be discussed further in the following section, and individual simulations will be highlighted.

\subsection{Capture Results}

In all the simulations run, no captured Trojan population was found to be significantly asymmetric between the L4 and L5 swarms. Figure \ref{L4L5Ratio} illustrates the number captured into each cloud by each simulation, and no points fall outside the 95\% confidence region. 

\begin{figure*}[t]
\centering
\includegraphics[width=1.0\textwidth]{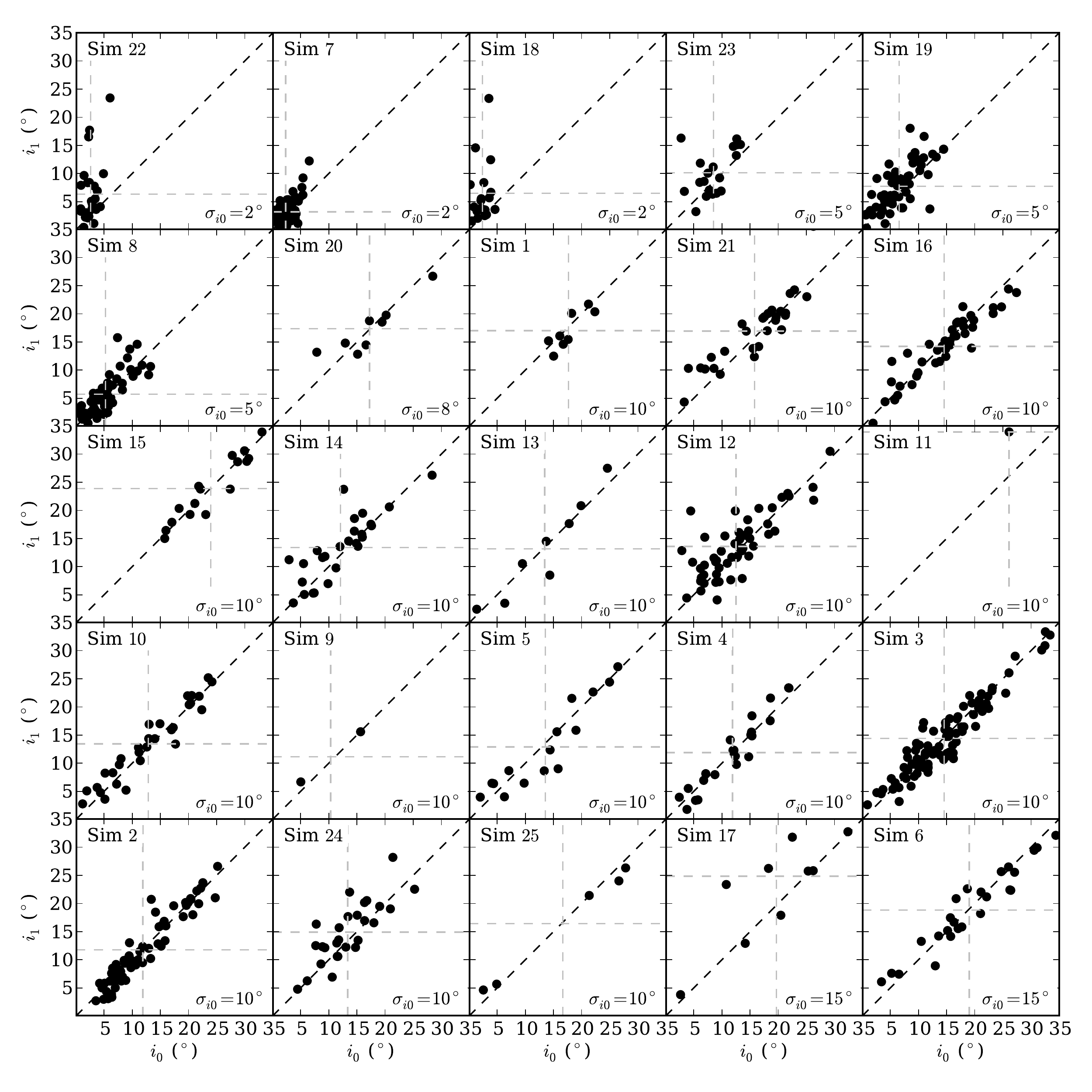}
\caption{Initial and final inclinations of disk particles that became Trojans. Simulation index is indicated, as is the initial disk width $\sigma_{i0}$. Dashed gray lines indicate the means of initial and final inclinations of the captured populations. Black dashed line indicates $i_0 = i_1$.}
\label{all_sim_results}
\end{figure*}

\begin{figure}[t]
\centering
\includegraphics[width=0.45\textwidth]{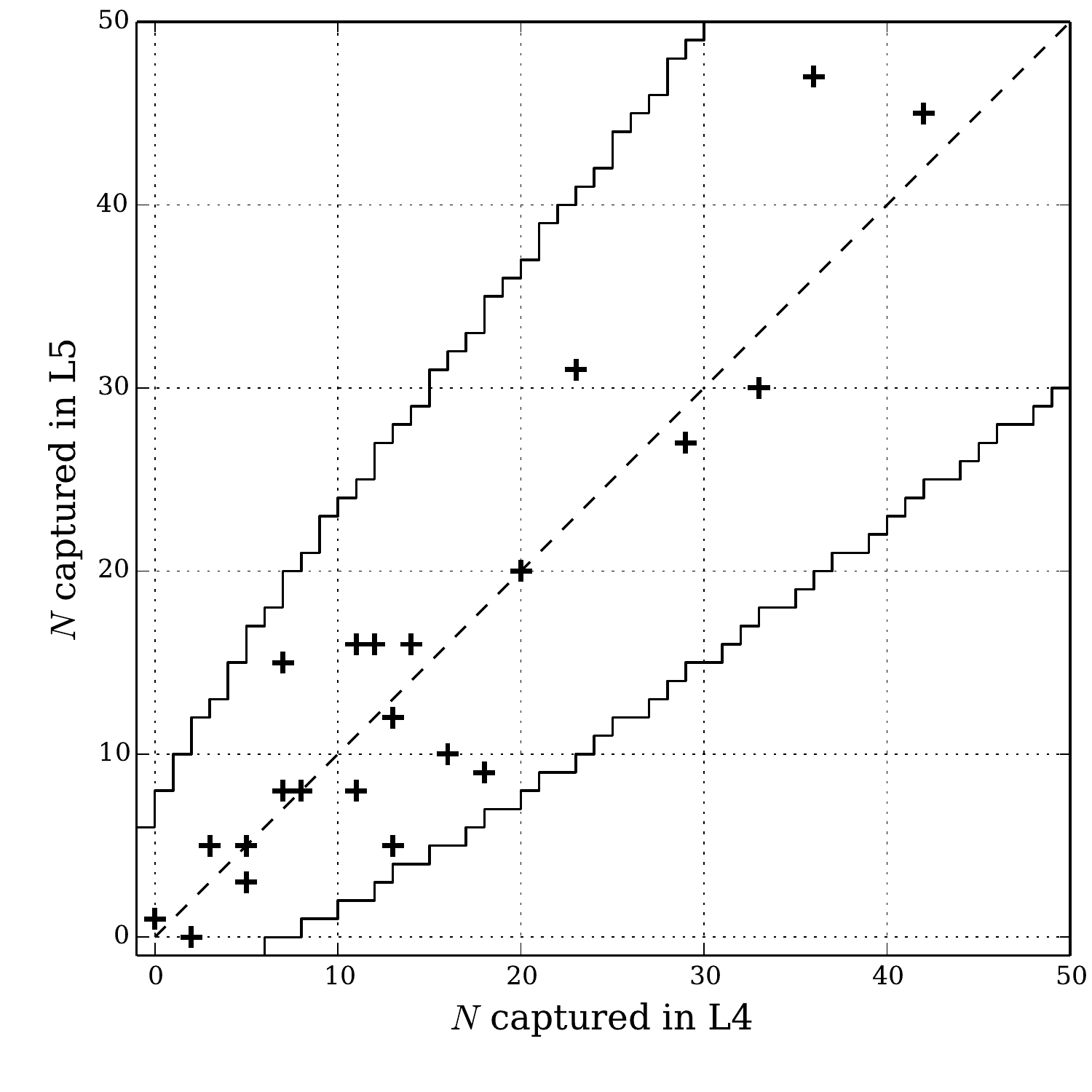}
\caption{Number of captures into each Trojan cloud by each migration simulation. Contours indicate symmetry deviation significant at the 95\% level for a given sample size according to a binomial test. No capture simulation produced a significantly asymmetric population.}
\label{L4L5Ratio}
\end{figure}

\begin{figure}[t]
\centering
\includegraphics[width=0.45\textwidth]{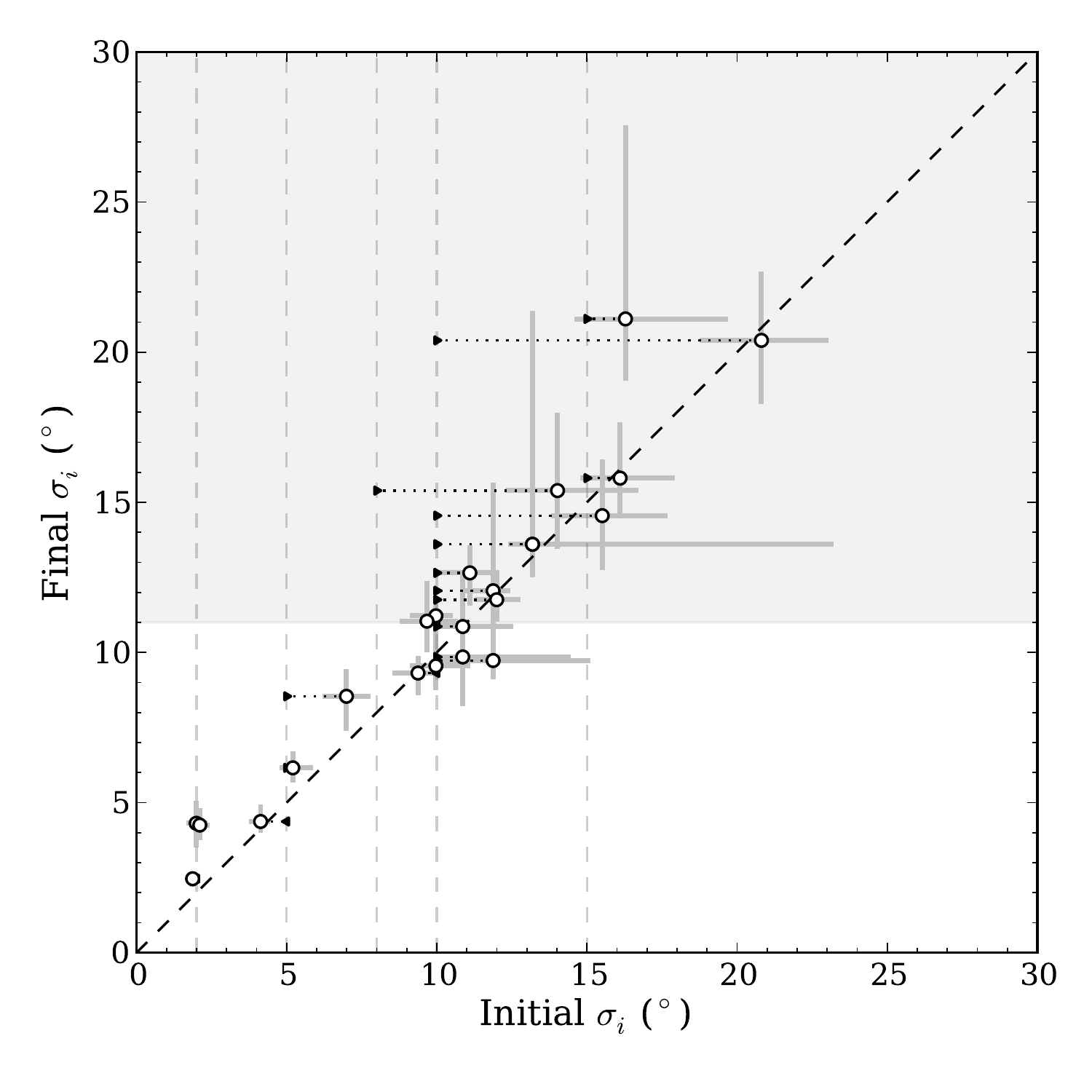}
\caption{Initial and final inclination widths of captured Neptune Trojans, the disk particles that became Trojans, and the original disks. Triangles indicate the initial disk width $\sigma_{i0}$, while circles indicate widths of Brown's distributions fit to the initial inclinations of \emph{just those particles that would go on to become Trojans,} and vertical location indicates the final width of the captured Trojans.}
\label{sigmain}
\end{figure}

Inclinations of captured Trojan particles were found to change little for objects on initially inclined orbits. In general, most simulations resulted in the preferential capture of particles with higher initial inclinations, resulting in final widths higher than initial disk widths due in large part not to stirring or scattering of particles through the capture process, but rather due to this preference for capturing \emph{initially} inclined objects. This produces a trend whereby the captured Trojan width is nearly identical to the initial width of \emph{just those particles that would go in to become Trojans}; see Figure \ref{sigmain}.  This indicates that given the rates of Neptune migration explored, the planetesimal disk must have had a significant component of high-inclination objects \emph{before} Neptune's arrival in order to produce the broad inclination widths that we see today; indeed, initial widths of $8^\circ - 10^\circ$ are required to produce widths consistent with the lower limit of the 95\% confidence interval for the Neptune Trojan inclination width.

\begin{figure}[h]
\centering
\includegraphics[width=0.45\textwidth]{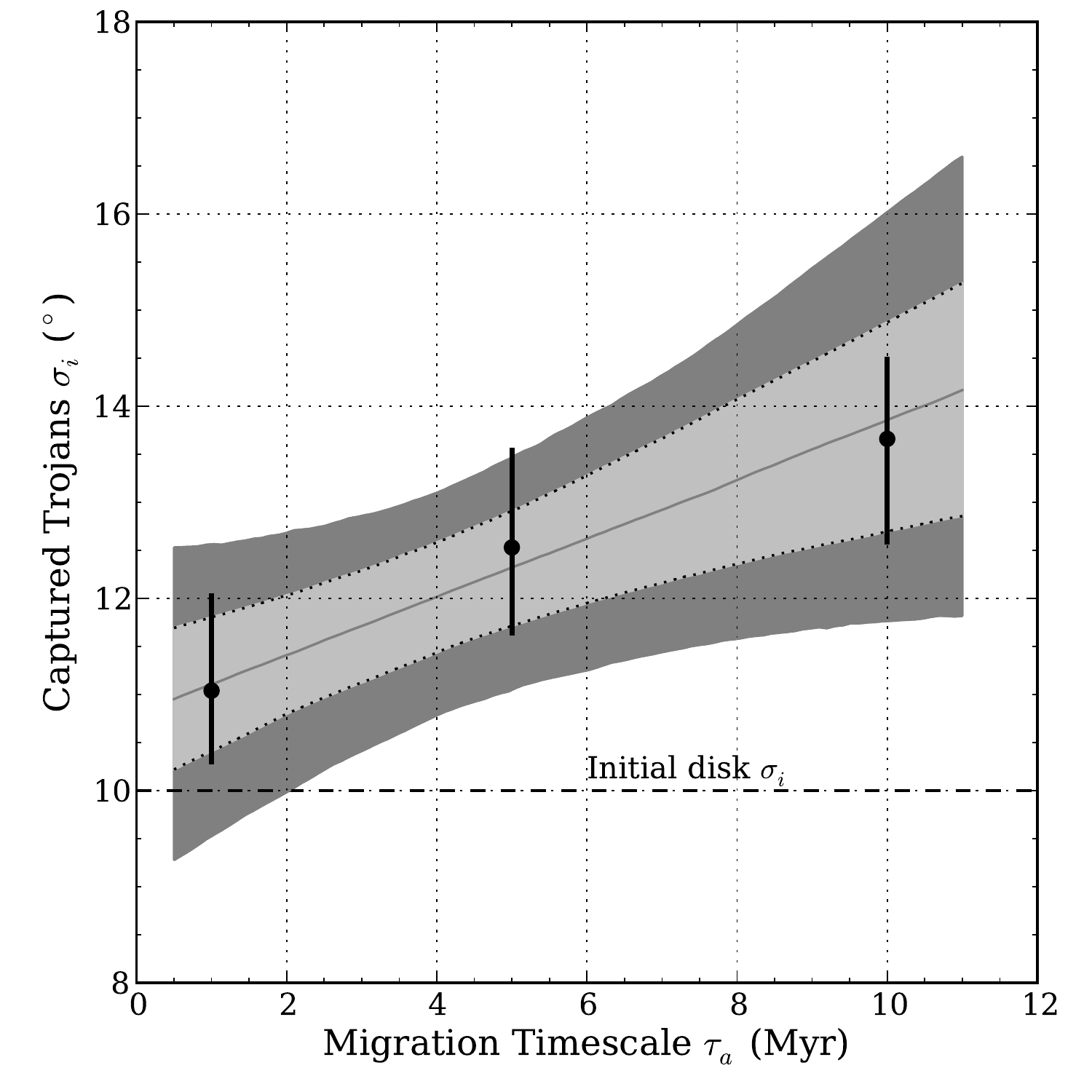}
\caption{ Three Neptune Trojan capture simulations with identical initial disks ($\sigma_{i0}=10^\circ$) and initial giant planet configurations, where only the migration timescale $\tau_a$ was varied from $10^6$ to $10^7$ years. Points indicate the individual simulations and estimates on the uncertainty of captured trojan $\sigma_i$; contours indicate acceptable regions spanned by linear models of $\sigma_i$ vs. $\tau_a$ (light gray indicates 68\% confidence contour, dark gray indicates 95\% confidence contour.)}
\label{wvt}
\end{figure}

In two integrations (15 and 20), the captured Trojan width is substantially higher than the initial disk width; in these cases, Neptune's initial eccentricity was set to 0.2, and the Uranus:Neptune period ratio was allowed to sweep through a moderate range of values. Other simulations with the same initial giant planet architecture, but colder initial disks ($\sigma_i = 2^\circ-5^\circ$) did not produce any captured Trojans and are not listed in Table 3. However, as in most other simulations, in these two integrations the captured Trojans were not heated by the capture process; on average, they ended up with nearly identical final inclinations and initial inclinations. This indicates that the capture process did not heat the Trojans; instead, the capture process \emph{preferred} \emph{pre}-heated objects. As such, in order to produce large observed widths, a primordial disk must already be stirred to large widths to have sufficient high-inclination objects to populate the high inclination Trojan orbits; if the disk is too cold, and few objects are of sufficiently high inclination to be captured, the mean Neptune Trojan capture probability plummets. In both of these simulations, the lowest initial inclination of a captured object was $8^\circ$. Thus, while these integrations seem to suggest that very broad inclination distributions can be generated from initially modest disk widths, they \emph{still} require that the initial disk have a significant population of high-inclination objects in order to populate the high-inclination Trojan orbits.

In several integrations with cold ($\sigma_i = 2^\circ$) primordial disks (eg., 18 and 22), several particles are visible in Figure \ref{all_sim_results} that attained captured inclinations substantially higher than their primordial values. However, the frequency of this occurrence in these simulations was low enough that the mean inclination was not substantially increased.

Longer migration timescales tended to produce larger captured inclination widths, as illustrated in Figure \ref{wvt}. Again, this increase in width seems to be largely due to an increasing preference for capturing inclined objects, instead of from stirring during the capture process. The increase in captured Trojan inclination width is relatively slow, increasing by only $\sim$20\% over an order of magnitude change in migration timescale. Recent arguments that the migration of Neptune may have occurred on timescales much longer than those explored here (Morbidelli et al. 2014) may provide an alternative explanation. If the relatively slow increase in captured population width with increasing migration timescale changes slope beyond the regimes explored here, the long migration timescales suggested in Morbidelli et al. (2014) could produce broad captured Trojan inclination distributions from a cold initial disk. Naive extrapolation from the simulations performed here suggest that capture efficiencies in such a scenario would be low, and may require problematically-massive initial disks if reduced by several orders; however, as previously noted, the capture efficiencies determined by the simulations presented here could be very sensitive to the assumptions of the primordial disk's radial structure and extent. Investing in simulations of capture simulations with much longer migration timescales than those considered here would be worthwhile.

\subsubsection{Integration Length}

Clones of two integrations (3 and 12) were extended beyond the nominal 100 Myrs, and continued out to 1 Gyr (4 and 13) to explore the long-term evolution of the captured populations. In both cases, the captured population decayed by factors of a few, and the nominal inclination distribution widths decreased marginally but the change was not statistically significant. In both cases, the nominal libration amplitudes decreased; for integration 4, the libration amplitude width decrease was statistically significant, dropping from $\sim19^\circ$ to $\sim13^\circ$. For integration 13, the nominal libration amplitude width decrease was smaller and not statistically significant given the captured Trojan sample sizes.

\section{Summary}

The Neptune Trojans have a high mean inclination; this work has demonstrated that if the population's inclination distribution is well-modeled by a truncated Brown's distribution like other resonant minor planet populations, then the width of that distribution must be greater than $11^\circ$ under conservative assumptions. By simulating the capture of Neptune Trojans from a dynamically-excited disk by a migrating, eccentric Neptune, I have shown that the inclination widths generally do not change drastically between the initial disk and the captured Trojan population; what change there is does not come in large part from stirring during the capture process, but rather from preferential capture of inclined objects. This indicates that objects with excited orbits today were likely excited prior to Neptune's arrival. 

A mechanism that might be responsible for this excitation is not yet clear. Migration of planets driven by interactions with excited disks has not yet been explored in great detail. The evidence presented in this work indicates that if Neptune migrated quickly, the disk Neptune migrated into must have been heated prior to Neptune's arrival; further work investigating precisely how Neptune's migration would have proceeded when driven by such a disk is merited.

\section{Acknowledgements}

The computations in this paper were run on the Odyssey cluster supported by the FAS Division of Science, Research Computing Group at Harvard University. I thank Matthew Holman, Rebekah Dawson, and Konstantin Batygin for their input regarding computational techniques and processes involved in planet migration and resonant capture, and Wesley Fraser for useful discussions on statistical techniques for minor planet population synthesis. Much of this manuscript was written in the \emph{3 Little Figs} cafe in Somerville, Massachusetts, and I thank the staff and owners for their coffee, lavender biscuits, and patience.


\appendix

\section{Means of truncated distributions}
\subsection{Rayleigh distribution}\label{RayMeanAppendix}

The PDF of the Rayleigh distribution is given by

\begin{equation}
f(i|\sigma) = (i/\sigma^2) e^{-\frac{1}{2}(i/\sigma)^2}.
\end{equation}

If this distribution is allowed to be continuous over all positive reals, then the mean is given by $\langle i \rangle = \sqrt{\frac{\pi}{2}} \sigma$. However, in the case where there is a finite upper limit $i_t$ applied, the mean is not as straightforward. The mean of such a truncated distribution with known PDF $f(i|\theta)$ and CDF $F(i|\theta)$ (in the case where the PDF and CDF are defined for arguments greater than 0) is given by

\begin{equation}
\langle i \rangle = \frac{ \int_0^{i_t} i f(i|\theta) di }{F(i_t|\theta)}
\end{equation}

In the case of a Rayleigh distribution, this evaluates to

\begin{eqnarray*}\label{RayMean}
\langle i \rangle &=& \frac{  \int_0^{i_t} (i^2/\sigma^2) e^{-\frac{1}{2}(i/\sigma)^2} di }{ 1 - e^{\frac{1}{2}(i_t/\sigma)^2} }
\\ &=& \frac{ \sigma \sqrt{\frac{\pi}{2}}\mbox{erf}\left( i_t / \sqrt{2}\sigma \right) -i_t e^{-\frac{1}{2}(i_t/\sigma)^2}}{ 1 - e^{-\frac{1}{2}(i_t/\sigma)^2}  }.
\end{eqnarray*}

Since as $\sigma$ grows large, the truncated distribution approaches $f(i) \propto i$, the mean of this truncated Rayleigh distribution has the asymptotic value of $\langle i \rangle \rightarrow \frac{2i_t}{3}$ as $\sigma \rightarrow \infty$. 

Eqn. \ref{RayMean} is a transcendental equation, so in order to propose a $\langle i \rangle$ as done in the text, Eqn. \ref{RayMean} must be numerically solved for $\sigma$. In my implementation, this was accomplished with the \textit{SciPy} implementation of Brent's method.

As long as the truncation value $i_t$ is greater than $\sigma$, then the mode of the truncated Rayleigh distribution remains $\sigma$. The median of the truncated Rayleigh distribution is trivial to compute, but I include it for completeness here:

\begin{equation}
i_{Me} = \sigma \sqrt{ -2 \ln \left( \frac{1}{2} ( e^{ - \frac{1}{2} ( i_t / \sigma)^2 } + 1 )  \right) }
\end{equation}

\noindent which, as $i_t \rightarrow \infty$ approaches the limit of the un-truncated distribution of $i_{Me} = \sigma \sqrt{ \ln(4) } $ .

\subsection{Brown's distribution}\label{BrownsMeanAppendix}

The Brown's distribution (Brown 2001) is colloquially given by

\begin{equation}
p(i)\propto \sin(i)e^{-\frac{1}{2}(i/\sigma)^2}
\end{equation}

For this work, we will assume that it applies only to the range $0\leq i \leq \frac{\pi}{2}$ (neglecting retrograde orbits). This means that, by nature, the Brown's distribution does not have a simple PDF, as it is already truncated. Additionally, since integrals of the form

\begin{equation}
\int \sin(i)e^{-\frac{1}{2}(i/\sigma)^2}di
\end{equation}

\noindent do not evaluate to a convenient form, determining the normalization of a PDF or CDF is not trivial. In order to determine the mean value, then, we must evaluate

\begin{equation}
\langle i \rangle = \frac{ \int_0^{i_t} i \sin(i)e^{-\frac{1}{2}(i/\sigma)^2}di }{ \int_0^{i_t} \sin(i)e^{-\frac{1}{2}(i/\sigma)^2}di }.
\end{equation}

To simplify this prospect, we use the Taylor expansion of $\sin(i)$ to make these integrals more similar to those we computed for the Rayleigh distribution. For truncations up to $\sim\pi/2$, the Taylor series up to order 5 is sufficient for our purposes.

\begin{eqnarray*}\label{BrownMeanApprox}
\langle i \rangle &\simeq& \frac{ \int_0^{i_t} (i^2 -\frac{1}{6}i^4 + \frac{1}{120}i^6) e^{-\frac{1}{2}(i/\sigma)^2}di }{ \int_0^{i_t} (i -\frac{1}{6}i^3 + \frac{1}{120}i^5) e^{-\frac{1}{2}(i/\sigma)^2}di } 
\\&=&  \frac{ A_1\mbox{erf}\left( i_t / \sqrt{2}\sigma \right) + A_2 e^{-\frac{1}{2}(i_t/\sigma)^2 }  }{ A_3 + A_4 e^{-\frac{1}{2}(i_t/\sigma)^2 } }.
\end{eqnarray*}

\noindent where

\begin{eqnarray*}
A_1 &=& 15\sigma \sqrt{2\pi}(8 - 4 \sigma^2 + \sigma^4)   \\
A_2 &=&  -2i_t (120 + 15 \sigma^4 -20i_t^2 + i_t^4 + 5 \sigma^2(i_t^2 - 12) )  \\
A_3 &=&  16(15 - 5 \sigma^2 + \sigma^4) \\
A_4 &=& - 2 ( 8 ( 15 - 5 \sigma^2 + \sigma^4) + 4 i_t^2 (\sigma^2 - 5) + i_t^4 ).
\end{eqnarray*}

Given that as $\sigma$ grows large, the truncated distribution approaches $f(i) \propto \sin(i)$, the mean of a truncated Brown's distribution has the asymptotic value of $\langle i \rangle \rightarrow \frac{\sin(i_t) - i_t\cos(i_t)}{1-\cos(i_t)}$ as $\sigma \rightarrow \infty$. For the approximation given by Eqn. \ref{BrownMeanApprox} the astymptotic limit is $(6i_t/7)(280 - 28i_t^2 +i_t^4)/(360-30i_t^2+i_t^4)$. For truncations up to $i_t \leq \pi/2$, these asymptotic values agree to within 0.05\%.

As with the truncated Rayleigh distribution, this must be solved numerically for $\sigma$ given a desired $\langle i \rangle$, and this was accomplished with the same numerical method described previously.

\section{Generating a sample of synthetic Neptune Trojan orbits}

Given a proposed parametric model for orbit distributions, I generate a synthetic sample of Neptune Trojan orbits and observational circumstances in the following way. First, inclinations $i$, libration amplitudes $L_{11}$, and eccentricities $e$ are drawn from their respective proposed distributions (semi-major axes are assumed to be identical). Libration phases $\psi_{11}$ are uniformly sampled over $[0,2\pi)$, and the current resonant arguments are generated by assuming sinusoidal libration behavior: $\phi_{11} - \langle \phi_{11} \rangle = | \lambdabar - \lambdabar_N |  - \langle \phi_{11} \rangle = L_{11} \sin( \psi_{11} )$. 

\begin{figure}[t]
\centering
\includegraphics[width=0.5\textwidth]{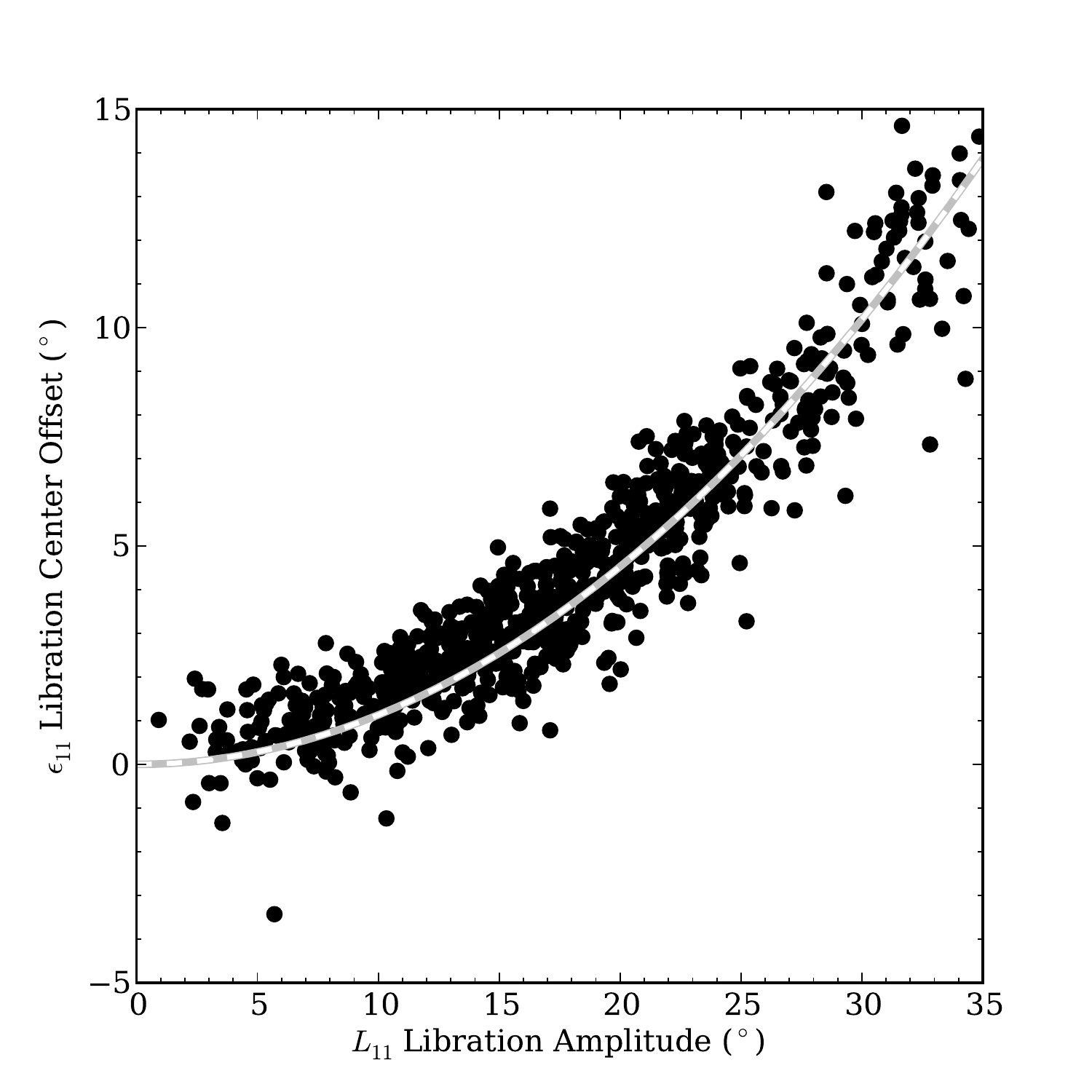}
\caption{ Offsets from ideal $\pm60^\circ$ libration centers found for synthetic Neptune Trojans. Dashed line indicates trend of $\epsilon_{11} = (L_{11} / 9.4^\circ )^2$, the adopted model used to produce synthetic Neptune Trojan observation circumstances.}
\label{LibOffset}
\end{figure}

Since I am defining the distribution with respect to the centers of the L4 and L5 clouds, I set $\langle \phi_{11} \rangle = 0^\circ + \epsilon_{11}$ instead of $\pm60^\circ$, where $\epsilon_{11}$ is a small, positive angle derived from the empirical libration properties of the synthetic Neptune Trojans generated by the capture simulations in this work. This offset was found to depend on $L_{11}$, $e$, and (weakly) on $i$. True majority of the variation in mean libration center was found to be well-represented by $\epsilon_{11} = (L_{11} / 9.4^\circ )^2$; see Fig. \ref{LibOffset}. Longitudes of ascending node $\Omega$ are uniformly sampled over $[0,2\pi)$. 

Instead of proposing an argument of pericenter and mean anomaly, then solving Kepler's equation, I use a small-eccentricity approximation to probabilistically determine an offset from the object's mean longitude caused by the object's eccentricity (the epicyclic component of motion). For eccentricities $\lesssim 0.1$, a sample of approximate offsets between true anomaly and mean anomaly can be drawn from $\Delta = 2 e \sin(U)$ where $U$ is a uniform random variate drawn over $[-\pi, \pi)$, and $\Delta$ is in Radians. 

Now it is possible to construct the cartesian coordinates of the Trojans with respect to the libration center:

\begin{eqnarray}
x &=& \cos( \phi_{11} + \Delta-\Omega) \cos(\Omega) - \sin( \phi_{11} + \Delta-\Omega) \cos( i ) \sin(\Omega)\\
y &=& \cos( \phi_{11} + \Delta-\Omega) \sin(\Omega) + \sin( \phi_{11} + \Delta -\Omega) \cos( i ) \cos(\Omega)\\
z &=& \sin( \phi_{11} + \Delta-\Omega) \sin(i) 
\end{eqnarray}

The two other observational parameters needed are the longitudinal angle between the Trojan and the mean Trojan libration center, $\lambda' = || \lambda - \lambda_N | - 60^\circ |$, and the absolute value of the Heliocentric latitude $\beta' = | \beta |$. From the cartesian coordinates,

\begin{eqnarray}
\lambda' &=& | \mbox{atan}_2(y, x) |\\
\beta' &=& | \mbox{asin}(z) |.
\end{eqnarray}

Now that $i$, $\lambda'$, and $\beta'$ are available, they are passed through their respective coverage functions; these functions represent the probability that a given object will be observed given its $i$, $\lambda'$, or $\beta'$. ``Synthetic observed'' Trojans are randomly sampled given the product of their coverage functions. The biased, ``synthetic observed'' sample of ($i,L_{11},e,\lambda', \beta'$) is returned to be compared directly to the real Neptune Trojan properties.

This approach implicitly assumes that libration is approximately sinusoidal, eccentricities are low, and the L4 and L5 clouds are identical.

\end{document}